\documentclass[prd,aps,amsfonts,preprint,floats,tightenlines,floatfix,
nofootinbib,showpacs,showkeys]{revtex4}
\usepackage{exscale}                  
\usepackage[intlimits]{amsmath}       
\usepackage{amsfonts}
\usepackage{amssymb,amscd}
\usepackage{epsfig}
\usepackage{wrapfig}
\newcommand{\beqa}{\begin{eqnarray}}
\newcommand{\eeqa}{\end{eqnarray}}
\newcommand{\beq}{\begin{equation}}
\newcommand{\eeq}{\end{equation}}
\newcommand{\pslash}{p\!\!\!/\,}
\newcommand{\qslash}{q\!\!\!/\,}

\begin{document}

\date{\today} \title{
 \hspace*{\fill}{\small\sf UNITU-THEP-6/04}\\
 \hspace*{\fill}{\small\sf IPPP/04/25 \hspace*{5mm}DCPT/04/50}\\
 \hspace*{\fill}{\small\sf http://arXiv.org/abs/hep-ph/0407104}\\[4mm]

Dynamical Chiral Symmetry Breaking in Unquenched $\mbox{QED}_3$}

\author{C.~S.~Fischer,}
\affiliation{IPPP, University of Durham, Durham DH1 3LE, U.K.}

\author{R.~Alkofer, T.~Dahm,}
\affiliation{Institute for Theoretical Physics, U.\ of T\"ubingen, 
D-72070 T\"ubingen, Germany}

\author{and P.~Maris}
\affiliation{Dept. of Physics and Astronomy, U.\ of Pittsburgh, 
Pittsburgh, PA 15260, U.S.A.}

\begin{abstract}
We investigate dynamical chiral symmetry breaking in unquenched
$\mbox{QED}_3$ using the coupled set of Dyson--Schwinger equations for
the fermion and photon propagators.  For the fermion-photon
interaction we employ an ansatz which satisfies its
Ward--Green--Takahashi identity.  We present self-consistent
analytical solutions in the infrared as well as numerical results for
all momenta.  In Landau gauge, we find a phase transition at a
critical number of flavours of $N_f^{\mathrm crit} \approx 4$.  In the
chirally symmetric phase the infrared behaviour of the propagators is
described by power laws with interrelated exponents.  For $N_f=1$ and
$N_f=2$ we find small values for the chiral condensate in accordance
with bounds from recent lattice calculations.  We investigate the
Dyson--Schwinger equations in other linear covariant gauges as well.
A comparison of their solutions to the accordingly transformed Landau
gauge solutions shows that the quenched solutions are approximately
gauge covariant, but reveals a significant amount of violation of
gauge covariance for the unquenched solutions.
\end{abstract}

\maketitle

\section{Introduction}
Over the years, quantum electrodynamics in (2+1) dimensions
($\mbox{QED}_3$) has been studied for a variety of reasons.  On the
one hand it served as a laboratory for investigating nonperturbative
phenomena such as dynamical mass generation or confinement in a
comparatively simple framework devoid of the technical complications
of non-abelian gauge theories (for reviews see
Refs.~\cite{Roberts:dr,Alkofer:2000wg,Mavromatos:2003ss}).  On the
other hand $\mbox{QED}_3$ has regained recent interest due to possible
applications in condensed matter systems.  High-$T_c$ cuprate
superconductors possess an unconventional $d$-wave symmetry of the
pairing condensate.  Such a pairing gap has nodes at the electronic
Fermi surface at which the low energy dispersion becomes linear and
thus can be described as massless fermions.  Since the electronic
motion is mainly confined to the two-dimensional copper-oxygen planes
in these systems an effective low energy description of the cuprates
in terms of a quantum electrodynamics in two spatial dimensions with
two massless fermion flavours has been suggested
\cite{Herbut:2002yq,Franz:2002qy,Nersesyan}.  In this picture the
antiferromagnetically ordered insulating state of the cuprates would
correspond to a state of broken chiral symmetry.  For this reason
there would be considerable interest in a study of the chiral phase
transition as well as the infrared spectral properties of the fermion
propagator in both the chirally symmetric and in the ordered phase of
$\mbox{QED}_3$.

QED in (2+1) dimensions is a super-renormalisable theory and has an
intrinsic mass scale given by the dimensionful coupling constant
$\alpha = N_f \, e^2/8$.  With the help of the photon polarisation
$\Pi(p)$ a dimensionless running coupling $\bar{\alpha} =
\alpha/(p[1+\Pi(p)]) = N_f \, e^2/(8 p[1+\Pi(p)])$ can be defined
which separates the nonperturbative infrared momentum regime from the
perturbative ultraviolet behaviour \cite{Appelquist:1986fd}.  A
nonzero fermion mass would provide a second mass scale.  Various
studies of the Dyson--Schwinger equation (DSE) of the fermion
propagator suggest that, in the chiral limit, the interactions
generate a dynamical fermion mass $M(p^2)$ (at least for a small
number of fermion flavours), and that this generated mass scale
$M(p^2=0)$ is considerably smaller than the scale defined by the
coupling constant $\alpha$ \cite{Appelquist:1986fd,
Appelquist:1988sr,Pennington:1988jw,Atkinson:1989fp,
Pennington:1990bx,Burden:mg,Maris:1995ns,Gusynin:1995bb,Maris:1996zg}.

It is the smallness of this generated mass scale that poses problems
in lattice Monte-Carlo simulations of $\mbox{QED}_3$
\cite{Dagotto:1988id,Hands:1989mv,Hands:2002dv,Hands:2004bh}.  Finite
volume effects are large and the relevant signal to determine the
chiral phase transition, the dimensionless chiral condensate, is very
small.  Furthermore the presence of an infrared cutoff as such has
been shown to reduce the value of the critical number of flavours,
$N_f^{\mathrm crit}$ \cite{Gusynin:2003ww}.  Thus recent studies for
the number of flavours $N_f=2$ \cite{Hands:2002dv} and $N_f=4$
\cite{Hands:2002dv, Hands:2004bh} determined bounds on the chiral
condensate, but no definite value for $N_f^{\mathrm crit}$ could be
extracted.  A definite signal for chiral symmetry breaking was
obtained only for $N_f=1$ \cite{Hands:2004bh}.  Given these problems
it seems evident that a continuum method is needed to shed light on
the infrared properties of $\mbox{QED}_3$.

The DSEs of the propagators of $\mbox{QED}_3$ have long been
investigated employing various levels of approximation.  Early
investigations of the fermion DSE based on a large $N_f$ expansion
indicated chiral symmetry to be broken only if the number of flavours
$N_f$ is smaller than a critical value of $N_f^{\mathrm
crit}=32/\pi^2\approx 3.2$ to leading order in Landau gauge
\cite{Appelquist:1988sr} and $N_f^{\mathrm crit}=4/3(32/\pi^2)$
including next lo leading order corrections in a nonlinear gauge
\cite{Nash:1989xx}.  These results have been questioned in
Refs.~\cite{Pennington:1988jw,Atkinson:1989fp,Pennington:1990bx},
where it was argued that the $1/N_f$-expansion is not an appropriate
tool to address these nonperturbative phenomena.  Using a slightly
different truncation of the fermion DSE, it was found that chiral
symmetry is broken for all values of $N_f$, although the generated
mass scale is exponentially decreasing for increasing
$N_f$~\cite{Pennington:1990bx}.  Subsequent work on the {\em coupled}
DSEs for the fermion and the photon propagator, however, again found
chiral symmetry restoration for $N_f > N_f^{\mathrm crit}$, with a
value of $N_f^{\mathrm crit}$ between 3 and 4
\cite{Maris:1995ns,Maris:1996zg}.  All investigations so far are
either quenched or employ a fermion-photon interaction which
manifestly violates gauge symmetry.

Certainly, gauge invariance is a key property of a local quantum field
theory and has to play a vital role in these investigations.  Reliable
results from DSEs can only be expected if the fermion-photon vertex
respects local gauge symmetry.  A necessary (though not sufficient)
condition in this respect is given by the Ward--Green--Takahashi
identity (WGTI) \cite{Ward:xp}, which determines the 'longitudinal'
part of the fermion-photon interaction uniquely in terms of the
propagator functions \cite{Ball:ay}.  The remaining transverse part of
the vertex is unrestricted by the WGTI and has to be determined either
directly from the vertex DSE or modelled by a suitable ansatz.  As the
vertex DSE is considerably more complicated to solve than the ones for
the propagators all work up to now concentrated on the latter
strategy.  Constraints on the structure of the transverse part of the
vertex have been derived from gauge covariance
\cite{Burden:gy,Dong:jr} and multiplicative renormalisability
\cite{Curtis:1990zs}.  Also perturbative vertex corrections put
constraints on the transverse fermion-photon vertex
\cite{Kizilersu:1995iz,Bashir:1997qt,Bashir:rv,Bashir:2001vi,
Raya:2003fx,Bashir:2004hh} and proposals for nonperturbative
generalisations of these structures have been made
\cite{Bashir:1997qt,Bashir:rv,Bashir:2001vi,Raya:2003fx}.

A further important nonperturbative tool to assess the gauge
transformation properties of a given vertex ansatz are the
Landau--Khalatnikov--Fradkin transformations (LKFT) \cite{LK}.  These
transformation laws leave the DSEs and the WGTI form invariant and in
principle allow one to test whether a given ansatz for a vertex is
gauge covariant.  Such an investigation, however, is hampered by the
fact that the transformation law for the vertex is quite complicated.
Therefore an indirect strategy has been applied: one calculates the
propagator with a given vertex ansatz in the fermion and photon DSE in
various gauges and compares with the corresponding results from the
LKFT of the propagator \cite{Burden:gy,Raya:2003fx,Bashir:2002sp,
Bashir:2004yt}.  The success of this strategy has been limited by the
problem that the LKFT is formulated in coordinate space and the
necessary Fourier-transform can be carried out analytically only for
very special cases.

Our aim in this paper is to make progress in the direction of a gauge
covariant solution for the propagators of $\mbox{QED}_3$.  We
investigate the coupled set of DSEs for the fermion and photon
propagator employing a fermion-photon interaction which respects the
WGTI.  In addition, it contains a transverse part that has been shown
to respect the LKFT properties in the quenched massless case and is
thus a good starting point for unquenched $\mbox{QED}_3$.  With the
absence of the fermion mass scale in the symmetric phase, the
assumption of a power law behaviour of the dressing functions in the
infrared seems natural.  For the same reasons similar power laws have
been found for the ghost and gluon propagators in quenched and
unquenched $\mbox{QCD}_4$ \cite{vonSmekal:1997is,Atkinson:1998tu,
Fischer:2002hn,Fischer:2003rp}.  We will see how far we can get with
this assumption here, given the limitations of the chosen truncation
scheme.

The paper is organised as follows: In Sec.~\ref{sec2} we discuss the
DSEs for the fermion and photon propagator and define our truncation
for the fermion-photon interaction.  Furthermore, we recall the
ultraviolet behaviour of the propagators as known from methods such as
the $1/N_f$-expansion, the loop expansion, and the operator product
expansion.  In Sec.~\ref{IR-sec} we present an analytical
determination of the infrared behaviour of the coupled system of
fermion and photon equations in the symmetric phase.  We show that,
within the limits of our truncation scheme, the infrared behaviour of
the propagators in Landau gauge is given by simple power laws,
confirming a longstanding conjecture from perturbative arguments
\cite{Atkinson:1989fp,Pennington:1990bx,Appelquist:1981sf}.  We also
calculate the critical number of flavours $N_f^{\mathrm crit}$ for
chiral symmetry breaking using these analytic solutions in the
infrared, and determine the behaviour of the fermion scalar
self-energy close to $N_f^{\mathrm crit}$ for a simplified version of
our fermion-photon vertex.  Furthermore we investigate the gauge
dependence of these power law solutions.

Numerical results in the broken and symmetric phases are presented in
Sec.~\ref{secIII}.  In passing we re-analyse the quenched fermion and
photon DSEs and demonstrate that the Curtis--Pennington vertex
resolves an inconsistency in determining the chiral condensate from
the fermion propagator, noted in Ref.~\cite{Burden:mg}.  We then
proceed to solve the unquenched system of photon and fermion DSEs
employing various ans\"atze for the fermion-photon vertex.  No
further approximations are made.  Our results in Landau gauge nicely
reproduce the analytical results in the ultraviolet as well as in the
infrared momentum regime.  With our most elaborate vertex ansatz the
critical number of flavours is $N_f^{\mathrm crit} \approx 4$.  The
order parameter of the phase transition, the dimensionless chiral
condensate $(-\langle \bar{\Psi}\Psi\rangle)/e^4$ is very small, of
the order of $10^{-3}$ even in the quenched limit, and decreases
exponentially as one approaches the phase transition.  For $N_f=1$ and
$N_f=2$ we find values of the condensate in agreement with the lattice
bounds \cite{Hands:2002dv,Hands:2004bh}.  We conclude with a
discussion of our results in Sec.~\ref{summary}.

\section{\label{sec2} 
The Dyson--Schwinger equations in $\mbox{QED}_3$}
We consider $\mbox{QED}_3$ with a four-component spinor representation
for the Dirac algebra and $N_f$ fermions.  This allows a definition of
chiral symmetry similar to the cases of $\mbox{QED}_4$ and
$\mbox{QCD}_4$.  With massless fermions, the Lagrangian has a
$U(2N_f)$ ``chiral'' symmetry, which is broken to $SU(N_f) \times
SU(N_f) \times U(1) \times U(1)$ if the fermions become
massive\footnote{Note that in this formalism a fermion mass term is
even under parity.  It is also possible to formulate $\mbox{QED}_3$
with two-component spinors; however, in such a formulation there is no
``chiral'' symmetry, and a fermion mass terms breaks parity
\cite{Appelquist:qw}.}.  The order parameter for this symmetry
breaking is the chiral condensate.  The question is: is this chiral
symmetry broken dynamically?  We use the set of DSEs to investigate
this question.

\subsection{The fermion and photon propagators}

\begin{figure}
\centerline{ \epsfig{file=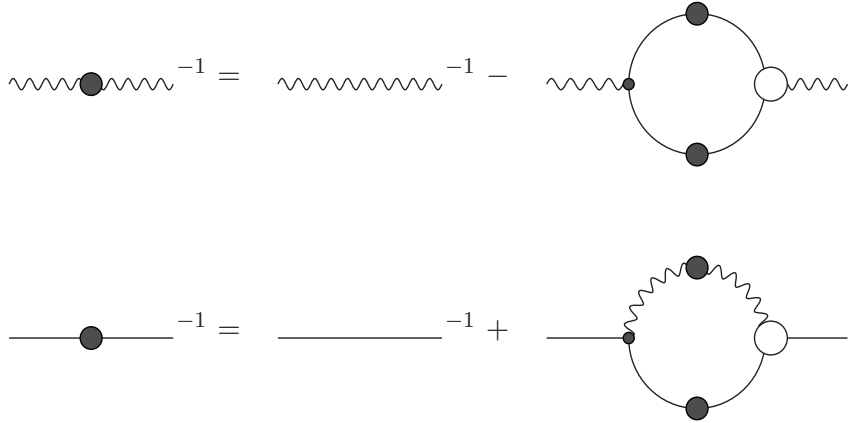,width=12cm,height=6cm} }
\caption{\label{fig00}
The Dyson--Schwinger equations of the photon and fermion propagators in
diagrammatic notation.}
\end{figure}

The DSEs for the photon and fermion propagators in
Euclidean space are given by
\beqa
D^{-1}_{\mu \nu}(p) &=& D^{-1}_{0,\mu \nu}(p) -
 Z_1 N_f e^2 \int \frac{d^3q}{(2 \pi)^3} \:\:
 \mbox{Tr}\left[ \gamma_\mu \,S(q)\, \Gamma_\nu(q,k) \,S(k) \right] \, ,
\label{fermionDSE}
\\
S^{-1}(p) &=& S^{-1}_0(p) +
 Z_1 e^2 \int \frac{d^3q}{(2\pi)^3} \: \gamma_\mu \: S(q) \:
 \Gamma_\nu(q,p)\: D_{\mu \nu}(k)  \, ,
\label{photonDSE}
\eeqa
with the momentum routing $k_\mu = q_\mu -p_\mu$.  A diagrammatic
notation of these equations is given in Fig.~\ref{fig00}.

The general form of the dressed fermion propagator
$S(p,\xi)$ and the photon propagator $D_{\mu \nu}(p,\xi)$ is given
by
\beqa
 S(p,\xi) &=& \frac{i \pslash A(p^2,\xi) + B(p^2,\xi)}
                   {p^2 A^2(p^2,\xi) + B^2(p^2,\xi)} \,,
\\
 D_{\mu \nu}(p, \xi) &=&
 \left(\delta_{\mu \nu} - \frac{p_\mu p_\nu}{p^2} \right) \,
 \frac{1}{p^2 (1+\Pi(p^2))} + \xi \frac{p_\mu p_\nu}{p^4} \, .
\label{photon}
\eeqa Here $\xi$ is the gauge parameter in linear covariant gauges,
with $\xi=0$ denoting Landau gauge.  The fermion functions $A$, $B$,
and $M$ depend on the gauge parameter $\xi$. On the other hand, the
vacuum polarisation $\Pi$ is independent of $\xi$.  Physical
quantities such as the fermion pole mass and the chiral condensate are
also independent of the gauge parameter.  In order to keep the
notation as clear as possible, we will treat all dependence of the
dressing functions on $\xi$ implicitly from now on.
 
The vertex normalisation constant $Z_1$ is related to the fermion wave
function normalisation $Z_2$ by a WGTI, $Z_1=Z_2$. Since
$\mbox{QED}_3$ is free of ultraviolet divergences, there is no need
for any renormalisation, though finite renormalisations of the fermion
and photon fields are possible and leave the physical content of the
theory invariant.  In our numerical procedure we set $A(\mu^2)=1$ at a
large normalisation point $\mu^2$ and determine $Z_2$
self-consistently.

\subsection{The fermion-photon vertex}
In general there are several possible strategies to choose an
appropriate approximation for the fermion-photon vertex
$\Gamma_\nu(q,p)$ in Eqs.~(\ref{fermionDSE}) and (\ref{photonDSE}).
The simplest option would be to replace the dressed vertex by the bare
vertex $\gamma_\nu$.  However, this violates, among other things,
gauge invariance and the renormalisation properties of the theory.  If
one wants to preserve these symmetries, one has to use a suitably
dressed vertex $\Gamma_\nu(q,p)$.

One way to dress the vertex would be to solve its corresponding DSE.
However, the vertex DSE contains an unknown four-point function, the
fermion-antifermion scattering kernel; one has to truncate the
infinite set of DSEs somewhere in order to obtain a tractable set of
equations, and this would only shift the problem up the hierarchy.
Furthermore, one faces the technical difficulties involved in solving
an integral equation in two independent momenta, i.e. in three
independent variables.  

A different strategy, which we adopt in this paper, is to employ an
ansatz for the vertex, which has to satisfy at least two requirements:
\begin{itemize}
\item[(a)] it must approach the perturbative form of the vertex for
large momenta;
\item[(b)] it must satisfy the WGTI
\beq
 i (q-p)_\nu \Gamma_\nu(p,q) = S^{-1}(p) - S^{-1}(q) \,.
\eeq 
\end{itemize}
Condition (a) reflects the fact that $\mbox{QED}_3$ is an
asymptotically free theory as explained in the introduction.  This
condition furthermore implicitly specifies the symmetry properties of
the vertex, i.e. its behaviour under charge conjugation and Lorenz
transformations.  Condition (b) is dictated by gauge invariance and
determines the longitudinal part of the vertex.  Furthermore, it
uniquely fixes the vertex when the two fermion momenta are equal
\beq
 \Gamma_\nu(p,p) = i\,\frac{\partial S^{-1}(p)}{\partial p_\nu} \;.
\eeq 
The two conditions (a) and (b) are necessary but not sufficient to
ensure gauge covariance of the propagators.  We will come back to this
point frequently later on.

Any vertex satisfying condition (b) leads to the following interesting
property of the fermion equation: With the explicit form, see
Eq.~(\ref{photon}), of the photon propagator in the fermion DSE, the
integral on the right hand side can be split into two pieces,
\beqa
 S^{-1}(p) &=& S^{-1}_0(p) + Z_1 \:e^2 \int \frac{d^3q}{(2\pi)^3} \: 
 \gamma_\mu \: S(q) \: \frac{1}{k^2\,(1+\Pi(k^2))} \:
 \left(\delta_{\mu \nu} - \frac{k_\mu k_\nu}{k^2} \right) 
 \: \Gamma_\nu(q,p) 
\nonumber\\ && \hbox{\hspace*{1.3cm}} 
 + Z_1 \:e^2 \: \xi\:\int \frac{d^3q}{(2\pi)^3} \: \gamma_\mu \: S(q) 
 \: \frac{k_\mu k_\nu}{k^4} \: \Gamma_\nu(q,p)\, ,
\label{DSE-exact}
\eeqa
with the momentum convention $k_\mu = q_\mu -p_\mu$ for the photon
momentum.  Due to the appearance of the longitudinal projection $k_\nu
\Gamma_\nu(q,p)$ in the second line of this equation, each vertex
truncation satisfying the WGTI treats this piece exactly.  This will
be important later on in our infrared analysis.

A suitable basis to construct a vertex ansatz satisfying the
requirements (a) and (b) has been given in Ref.~\cite{Ball:ay}.  It
consists of twelve tensor structures, which can be split up in a set
of four components, $\Gamma^{BC}_\nu$, completely determined by the
WGTI and eight transverse components, $\Gamma^T_\nu$
\beq
\Gamma_\nu(p,q) =  \Gamma^{BC}_\nu(p,q) + \Gamma^T_\nu(p,q) \, . 
\eeq
The WGTI is solved by the Ball--Chiu (BC) construction
\beqa
\Gamma_\nu^{BC}(p,q) &=& \frac{A(p^2)+A(q^2)}{2} \gamma_\nu + 
 i \frac{B(p^2)-B(q^2)}{p^2-q^2} (p+q)_\nu
\nonumber\\ && \hbox{}
 + \frac{A(p^2)-A(q^2)}{2(p^2-q^2)} (\pslash+\qslash)(p+q)_\nu \, . 
\label{BC}
\eeqa
The WGTI furthermore constrains the $\sigma_{\mu \nu} (p_\mu +
q_\mu)$-component of the vertex to be zero.  The eight transverse
components satisfy
\beq
 k_\nu \Gamma^T_\nu(p,q) =  0,\hspace*{2cm} \Gamma^T_\nu(p,p) =  0  \,,
\eeq
and are otherwise constrained by condition (a). Much work has been
invested to determine $\Gamma^T_\nu(q,p)$ in the perturbative region
\cite{Curtis:1990zs,Kizilersu:1995iz,Bashir:1997qt,Bashir:rv,
Bashir:2001vi,Raya:2003fx,Bashir:2004hh,Davydychev:2000rt} to
constrain possible nonperturbative ans\"atze, but so far without
conclusive results.

A minimal ansatz for $\Gamma^T_\nu(q,p)$ ensuring multiplicative
renormalisability in four-dimen\-sional quenched QED has been given by
Curtis and Pennington \cite{Curtis:1990zs}
\beq
\Gamma^{T,CP}_\nu(p,q) = \frac{A(p^2)-A(q^2)}{2} 
  \:\frac{\left[(p^2-q^2)\gamma_\nu - 
  (\pslash-\qslash)(p+q)_\nu \right] \left(p^2+q^2\right)}
  {(p^2-q^2)^2+(M^2(p^2)+M^2(q^2))^2}  \,.
\label{CP}
\eeq Burden and Roberts \cite{Burden:gy} discovered another favourable
property of the Curtis--Pennington (CP) vertex $\Gamma^{CP}_\nu(q,p) =
\Gamma^{BC}_\nu(q,p) + \Gamma^{T,CP}_\nu(q,p)$, which holds in {\em
quenched massless} $\mbox{QED}$ in both three and four dimensions:
with the help of the LKFT \cite{LK} for the propagators they showed
that the CP-vertex indeed preserves gauge covariance in these special
cases.  More sophisticated ans\"atze for the transverse parts of
the vertex have been given in Refs.~\cite{Dong:jr,Bashir:1997qt,
Bashir:rv,Bashir:2001vi,Raya:2003fx}.  Contrary to early expectations,
it has been noted \cite{Bashir:2001vi} that $\Gamma^T_\nu(p,q)$ does
not vanish in Landau gauge and contains terms that have to be
explicitly dependent on the gauge parameter $\xi$.

\subsection{The truncated Dyson--Schwinger equations}
A numerical investigation of all these ans\"atze, though highly
desirable, is a formidable task.  Up to now the quenched fermion DSE
and the photon DSE of $\mbox{QED}_3$ have been studied employing a
bare fermion photon vertex as well as the BC construction
\cite{Burden:mg,Burden:gy} and the CP-vertex \cite{Bashir:2002dz}.
Partly unquenched calculations can be found in
Refs.~\cite{Appelquist:1986fd,Appelquist:1988sr,Pennington:1988jw,
Atkinson:1989fp, Pennington:1990bx,Walsh:1990} where the photon
propagator has been approximated by its $1/N_f$-expression.  Fully
unquenched dynamical solutions employing the bare as well as the first
part of the BC-vertex have been reported in
Refs.~\cite{Maris:1995ns,Maris:1996zg}. In our work we will extend
these investigations and solve the unquenched equations employing the
CP construction in the fermion DSE and the BC-vertex in the photon
equation.  The reason for this hybrid choice is the following: The
transverse term in the CP-vertex has been constructed for quenched
QED, i.e. its structure is adapted to the kinematical situation in the
fermion DSE and it is believed to approximate some parts of the real
fermion-photon vertex that are important in the fermion DSE.  In the
photon DSE, however, a different kinematical region of the vertex is
probed.  The $\Gamma^{T,CP}$-term leads to divergences here
\cite{Bloch:1995dd} and is thus not a good approximation to the
important parts of the real vertex in the photon DSE.

Substituting the CP-vertex into the fermion DSE and
taking appropriate traces we arrive at
\beqa
\lefteqn{ B(p^2) \; = \; Z_2 \: e^2 \int \frac{d^3q}{(2\pi)^3} 
  \frac{1}{k^2 \: (q^2 A^2(q^2) + B^2(q^2))} \times   }
\nonumber\\ && \Bigg\{ 
  \frac{1}{1+\Pi(k^2)} \Bigg[\frac{A(p^2)+A(q^2)}{2} 2 B(q^2) 
   + \left(A(p^2)-A(q^2)\right) \: B(q^2) \: \Omega(p^2,q^2)
\nonumber\\ &&\hbox{\hspace*{19mm}}
   + \left(\Delta A\:B(q^2)-\Delta B A(q^2)\right)
   \left(-\frac{k^2}{2}+(p^2+q^2)-\frac{(p^2-q^2)^2}{2k^2}\right)
   \Bigg] 
\nonumber\\ && \hbox{\hspace*{-1mm}}
   + \xi \;
   \Bigg[\frac{A(p^2)+A(q^2)}{2} B(q^2) 
   + \Delta A\:B(q^2) \frac{(p^2-q^2)^2}{2k^2} - \Delta B\:A(q^2) 
   \left( \frac{(p^2-q^2)^2}{2k^2} + \frac{q^2-p^2}{2} \right) 
   \Bigg] \Bigg\}
\nonumber\\ \label{B-eq}  
\eeqa
\beqa
\lefteqn{ A(p^2) \;=\; Z_2 +  Z_2 \:e^2  \int \frac{d^3q}{(2\pi)^3} 
  \frac{A(q^2)}{p^2 \: k^2 \: (q^2 A^2(q^2) + B^2(q^2))} \Bigg\{ 
 \frac{1}{1+\Pi(k^2)}  \times }
\nonumber\\ && 
  \Bigg[ \frac{A(p^2)+A(q^2)}{2} 
                 \left(  -\frac{k^2}{2} + \frac{(p^2-q^2)^2}{2k^2} \right) 
  + \left(A(p^2)-A(q^2)\right) \: \Omega(p^2,q^2) 
    \left(\frac{p^2+q^2-k^2}{2}\right)
\nonumber\\ &&\hbox{\hspace*{19mm}}
  - \left(\Delta A \frac{p^2+q^2}{2}+\Delta B M(q^2)\right)
    \left(-\frac{k^2}{2}+(p^2+q^2)-\frac{(p^2-q^2)^2}{2k^2}\right) \Bigg]
\nonumber\\ && \hbox{}
  + \xi \; \Bigg[\frac{A(p^2)+A(q^2)}{2} \left(\frac{p^2+q^2}{2} 
		 - \frac{(p^2-q^2)^2}{2k^2}\right)
\nonumber\\ &&\hbox{\hspace*{7mm}}
 + \Delta A \frac{(p^2-q^2)}{2} 
   \left(\frac{p^2-q^2}{2}+\frac{q^4-p^4}{2k^2}\right)
 - \Delta B\:M(q^2) 
   \left( \frac{(p^2-q^2)}{2} + \frac{(q^2-p^2)^2}{2k^2} \right) 
   \Bigg] \Bigg\} 
\nonumber\\ \label{A-eq}
\eeqa
Here we used the abbreviations 
\beqa
\Delta A &=& \frac{A(p^2)-A(q^2)}{p^2-q^2} \, ,\nonumber\\
\Delta B &=& \frac{B(p^2)-B(q^2)}{p^2-q^2} \, ,\nonumber\\
\Omega(p^2,q^2) &=& \frac{p^4-q^4}{(p^2-q^2)^2+(M^2(p^2)+M^2(q^2))^2} \, .
\nonumber
\eeqa
Furthermore we have used the WGTI $Z_1 = Z_2$ for the vertex
and wave function renormalisation constants.

In the photon equation we contract the Lorenz indices with the general
tensor \cite{Brown:1988bn,Fischer:2002eq}
\beq
{\mathcal P}^{(\zeta )}_{\mu\nu} (p) = \delta_{\mu\nu} - 
\zeta \frac{p_\mu p_\nu}{p^2}  \, . \label{tensor} 
\eeq
Inserting the BC-vertex in the general expression for the vacuum
polarisation, we obtain
\beqa
 \Pi(p^2) &=&-Z_2 \: e^2 N_f \int \frac{d^3q}{(2 \pi)^3} 
\; \frac{1}{q^2 A^2(q^2) + B^2(q^2)} \; \frac{1}{k^2 A^2(k^2) + B^2(k^2)} 
 \times \nonumber\\ && 
\Bigg\{\frac{A(q^2)+A(k^2)}{2} \left(W_1(p^2,q^2,k^2) A(q^2)A(k^2) 
   + W_2(p^2,q^2,k^2) B(q^2)B(k^2)\right) 
\nonumber\\ && \hbox{}
 + \frac{A(q^2)-A(k^2)}{2(q^2-k^2)} \left(W_3(p^2,q^2,k^2) A(q^2)A(k^2) 
 + W_4(p^2,q^2,k^2) B(q^2)B(k^2) \right) 
\nonumber\\ && \hbox{}
 + \frac{B(q^2)-B(k^2)}{q^2-k^2} \left(W_5(p^2,q^2,k^2) A(q^2)B(k^2) 
 + W_6(p^2,q^2,k^2) B(q^2)A(k^2) \right) \Bigg\} \,. 
\nonumber\\  
\label{quarkloop}
\eeqa
The general form of the kernels $W_i$ as well as a discussion of the
(weak) dependence of the photon equation on the parameter $\zeta$ of
the projector (\ref{tensor}) can be found in
Appendix~\ref{app1}.  Here we give the kernels for the special value
$\zeta=3$, suggested in Ref.~\cite{Brown:1988bn} to avoid spurious
divergences
\beqa
W_1(p^2,q^2,k^2) &=&
  \frac{3 k^4}{ p^4} - k^2\left(\frac{2}{p^2} + \frac{6 q^2}{p^4}\right)
  -1 -\frac{2 q^2}{p^2}  + \frac{3 q^4}{p^4} \,, \label{w1}
\\
W_2(p^2,q^2,k^2) &=& 0 \,, 
\\	       
W_3(p^2,q^2,k^2) &=& 
  \frac{3 k^6}{ p^4} - k^4\left(\frac{4}{p^2} + \frac{3 q^2}{p^4}\right)
  +k^2 \left(1 -\frac{3 q^4}{p^4}  \right) 
  +q^2 - \frac{4q^4}{p^2} + \frac{3 q^6}{p^4} \,, 
\\ 
W_4(p^2,q^2,k^2) &=& 
  \frac{-6 k^4}{p^4} + k^2 \left(\frac{4}{p^2}+\frac{12 q^2}{p^4}  \right) 
  -2 + \frac{4q^2}{p^2} - \frac{6 q^4}{p^4}  \,,
\\ 
W_5(p^2,q^2,k^2) &=&  
  \frac{3 k^4}{p^4} -k^2\left( \frac{4}{p^2}+\frac{6 q^2}{p^4}\right)
  + 1 + \frac{3 q^4}{p^4}  \,,
\\ 
W_6(p^2,q^2,k^2) &=& 
  \frac{3 k^4}{p^4} - k^2 \left(\frac{6 q^2}{p^4} \right)
              + 1  - \frac{4 q^2}{p^2}+ \frac{3 q^4}{p^4} .\label{w6}
\eeqa

\subsection{\label{asymp-sec}
$1/N_f$-expansion and asymptotic behaviour}
Several previous studies of $\mbox{QED}_3$ have used the
$1/N_f$-expansion\footnote{This is equivalent to a perturbative
expansion for small $e^2$ while keeping $\alpha = N_f e^2/8$ fixed. As
$\mbox{QED}_3$ is an asymptotically free theory this expansion will
provide correct answers in the ultraviolet.}  to justify truncations
or to calculate the asymptotic behaviour of the dressing functions.
Our approach does not rely on this expansion, but it is interesting to
compare our results with the ones obtained in this way.  We therefore
shortly summarise the anticipated behaviour of the dressing functions
based on the $1/N_f$-expansion.

For the vacuum polarisation, one finds for $N_f$ massless fermion
flavours to leading order in a $1/N_f$-expansion \cite{Appelquist:1986fd}
\beq
\Pi(p^2) = \frac{N_f \; e^2}{8p} = \frac{\alpha}{p} \, , 
\label{UV_PH}
\eeq 
independently of the value for the gauge parameter $\xi$.  In the full
theory, this expression remains valid in the ultraviolet asymptotic
limit, as has been demonstrated in quenched approximation employing a
BC-vertex in the fermion loop \cite{Burden:1991uh}.  Our numerical
results show that this is also the case in the unquenched case.  In
the infrared we will find a modified power law for the photon.

The asymptotic behaviour of the vector self energy to two loop order 
in {\em quenched approximation} is given by \cite{Bashir:rv}
\beq
A(p^2 \rightarrow \infty) = 1+\frac{\xi e^2}{16 p} 
+ \frac{e^4 \xi^2}{64 \pi^2 p^2}
+ \frac{3 e^4}{64 \pi^2 p^2} \left(\frac{\pi^2}{4} 
- \frac{7}{3}\right) + O(\alpha^3) \label{UV_A}
\eeq
Note that to this order the vector dressing function $A$ receives
positive corrections in all gauges, i.e. $A(p^2) \downarrow 1$ for
$p^2 \to \infty$.  Furthermore, these corrections to its asymptotic
value $A=1$ behave like inverse powers of the momentum.

On the other hand, from an {\em unquenched} $1/N_f$-expansion in
Landau gauge, employing a bare vertex and the $1/N_f$ photon
propagator given in Eq.~(\ref{UV_PH}), one obtains for $p < \alpha$
\beq
 A(p^2) = 1+\frac{8}{3N_f \pi^2} \ln(p/\alpha) \, ,
\label{1/N_A}
\eeq 
up to terms that are regular for $p \to 0$.  Certainly this expression
cannot be valid in the ultraviolet region (as $A(p^2) \to 1$ for large
momenta), nor in the (far) infrared region, reflecting the
inconsistency of ignoring vertex corrections.  Nevertheless, it has
been argued \cite{Appelquist:1981sf,Atkinson:1989fp,
Pennington:1990bx} that it could be the first term in the build up of
an anomalous dimension
\beq 
 A(p^2) = \left(\frac{p^2}{\alpha^2}\right)^{\eta} \,,
\label{A-1/N} 
\eeq 
with 
\beq 
 \eta = \frac{4}{3\pi^2 N_f } \approx \frac{0.135}{N_f} \,,
\eeq
in the infrared region.  We will come back to this possibility in our
infrared analysis in the next section.  

Finally, the analysis of the asymptotic behaviour of the scalar
dressing function $B(p^2)$ in the chirally broken phase of massless
$\mbox{QED}_3$ is outlined in \cite{Burden:mg}.  Since we are
interested in dynamical chiral symmetry breaking, which is a purely
nonperturbative phenomenon, we cannot rely on the $1/N_f$-expansion to
obtain an expression for $B$.  Using the operator product expansion
however, one finds that asymptotically
\beq
 B(p^2 \rightarrow \infty) = 
 \frac{2+\xi}{4} \; \frac{\langle\bar{\Psi} \Psi \rangle}{p^2}  \,,
\label{UV_B}
\eeq
with $A(p^2 \rightarrow \infty) \rightarrow 1$.
Thus the chiral condensate $\langle\bar{\Psi} \Psi \rangle$ can be
obtained from the fermion propagator in two ways: on the one hand it
can be read off from the asymptotic behaviour of the B-function and on
the other hand it is given by the trace of the propagator in
coordinate space.  In Ref.~\cite{Burden:mg} slight deviations between
these two methods have been found.  We will demonstrate that these
deviations do not occur in our truncation.  The asymptotic form,
Eq.~(\ref{UV_B}), is reproduced to very good accuracy for a range of
values of the gauge parameter $\xi$ in our numerical analysis.

\section{\label{IR-sec}
Infrared asymptotic behaviour}
Unlike the situation in a perturbative analysis, where one has a
definite starting point to work out results order by order, an
analysis of the nonperturbative momentum regime of $\mbox{QED}_3$ 
is based solely on self-consistency and relies therefore on a physically
motivated ansatz to start with.  Inspired by the
conjecture of the previous subsection and guided by our numerical
analysis, our working hypothesis will be that at least in Landau gauge
the infrared behaviour of $\mbox{QED}_3$ in the symmetric phase is
given by power laws.  We will investigate this assumption employing
different vertex truncations and see how far we can get.  Finally we
will investigate whether our results are gauge covariant.

\subsection{\label{ir-landau}
Infrared analysis in Landau gauge}
\subsubsection{The symmetric phase}
For the following analysis we will not use the full CP or BC vertex
constructions but only the term proportional to $\gamma_\mu$, denoted
1BC,
\beq
 \Gamma_\mu^{1BC}(p,q,k) = \frac{A(p^2)+A(q^2)}{2} \gamma_\mu \, .
\eeq
This choice has the advantage that the equations are simplified
significantly and it already contains all qualitative features of the
solution employing the full CP/BC-vertex in the infrared region, as
will be demonstrated by our numerical calculations given in
Sec.~\ref{secIII}.  Furthermore, it has the merit that we can solve
the DSEs in the infrared analytically.  This vertex has been
considered before in Ref.~\cite{Maris:1996zg}, though no attempt
was made there to solve the DSEs analytically.  

The starting point of our investigation is a power law ansatz for
the vector dressing function
\beq
 A(p^2) = c\:p^{2\kappa} \label{power-law} \,,
\eeq
with the constant $c$ and the power $\kappa$ to be determined
self-consistently.  We expect this ansatz to be valid in the infrared,
i.e. in the momentum region $p \ll \alpha$; for $p > \alpha$ the
function $A(p^2)$ rapidly approaches its free form, $A(p^2)=1$.
The integrals on the right hand side of the DSEs are dominated by the
infrared contributions, coming from the region $p < \alpha$.  Thus one
can safely substitute the power law for $A(p^2)$ and cut off the
integrals at $p=\alpha$.  Alternatively, one can substitute the
power law over the entire momentum range in the integrals,
provided one keeps track of possible ultraviolet divergences.
After integration, the resulting power behaviour on the right hand
side of the equations then has to match the power law on the left hand
side.

Given the ansatz, Eq.~(\ref{power-law}), we first have to derive the
corresponding power law of the photon polarisation.  After
substituting the ansatz (\ref{power-law}) into the right hand side of
the photon equation, Eq.~(\ref{quarkloop}), the integral can be
carried out with the help of Eq.~(\ref{IR-int}).  We arrive at
\beqa
 \Pi_{1BC}(p^2) &=& Z_2 \: \frac{\alpha}{c} \:\frac{4}{\pi} \:  
  \frac{\Gamma(3/2-\kappa) \Gamma(1/2+\kappa)}{\Gamma(3-\kappa)  
 \Gamma(1+\kappa)} \:p^{-1-2\kappa} \, , 
 \label{ph-bc-ir}\\
  &=:& Z_2 \:\frac{\alpha}{c} \: w_{1BC}\:p^{-1-2\kappa} \, , 
\nonumber
\eeqa
where we have introduced the dimensionless function $w_{1BC}(\kappa)$ 
\beqa
  w_{1BC}(\kappa) &=:& \frac{4}{\pi} \:  
  \frac{\Gamma(3/2-\kappa) \Gamma(1/2+\kappa)}{\Gamma(3-\kappa)  
 \Gamma(1+\kappa)} \,.
 \label{ph-bc-kappa}
\eeqa 

For the analysis of the fermion DSE, Eq.~(\ref{A-eq}), we assume that
$\kappa > -1/2$, as $\kappa \le -1/2$ only admits the trivial solution
$A \equiv 1$, cf. Sec.~\ref{IR-nonlandau} below.  With $p \ll \alpha$
the photon dressing is given by 
\beq
\frac{1}{1+\Pi_{1BC}(p^2)} \approx 
    \frac{c}{\alpha w_{1BC}} \:\:p^{1+2\kappa} \,.
\eeq  
Together with the power law Eq.~(\ref{power-law}) we then obtain
\beqa
c\: p^{2\kappa} &=&  Z_2 +  \frac{c}{w_{1BC} \:N_f\: \pi^3} 
  \:\int d^3q \left\{ \frac{k^{-1+2\kappa}}{p^2 \,q^{2+2\kappa}} \:
  \frac{p^{2\kappa}+q^{2\kappa}}{2} 
   \left(  -\frac{k^2}{2} + \frac{(p^2-q^2)^2}{2k^2}
                        \right) \right\} \,.
\eeqa
The treatment of this type of equation has been discussed in detail in
Refs.~\cite{Lerche:2002ep,Zwanziger:2002ia} for the system of ghost
and gluon DSEs in $\mbox{QCD}_4$.  To proceed one has to distinguish
two cases, $\kappa<0$ and $\kappa>0$.  In the first case the left hand
side of the equation becomes singular for $p^2 \rightarrow 0$ and has
to be matched by a corresponding singularity in the integral on the
right hand side.  In this case the constant term, $Z_2$, stemming from
the bare propagator is suppressed and can simply be discarded.  (This
case is analogous to the gluon equation in $\mbox{QCD}_4$).  On the
other hand, if $\kappa>0$ the left hand side goes to zero.  The
renormalisation constant $Z_2$ thus has to be cancelled by a constant
term generated by the integral.  A straightforward way to deal with
this situation is to discard the constant term $Z_2$ and at the same
time to eliminate the constant term hiding in the integral by
employing dimensional regularisation.  Furthermore we have to
eliminate a spurious divergence introduced by employing the power law
ansatz over the whole momentum range.  We are then left with
\beqa
p^{2\kappa} &=& \frac{p^{2\kappa}}{w_{1BC} \:N_f\: \pi^2} 
  \left(\frac{1}{2\kappa(1-2\kappa)} 
  + \frac{\pi}{(3+2\kappa)} \frac{\Gamma(\kappa)\Gamma(1-\kappa)}
  {\Gamma(3/2-\kappa)\Gamma(1/2+\kappa)} \right)  \,.
\label{A-IR-1BC}	         
\eeqa
Note that the normalisation factor $Z_2$ as well as the coefficient
$c$ of the power law has been dropped out of the equation as 
expected.  The powers of momentum match on both sides of the equation
thus confirming that the power law is indeed a self-consistent
solution of the DSEs in the chirally symmetric phase.  Equations
(\ref{ph-bc-kappa}) and (\ref{A-IR-1BC}) together determine the
exponent $\kappa$ and therefore completely describe the behaviour of
the photon and fermion propagators in the infrared in the given
truncation scheme.

 From our analysis we find a possible explanation, why the authors of
Refs.~\cite{Atkinson:1989fp,Pennington:1990bx,Walsh:1990}
did not find a phase transition in their truncation scheme: as the
feedback from the function $A$ onto the vacuum polarisation is not
taken into account in their approach, i.e. $\Pi(p^2) \sim 1/p$, the
right hand side of the DSE for $A$ is proportional to $p^0$, which
only matches the left hand side iff the $A$-function becomes a (trivial)
constant in the infrared.  Thus there is no self-consistent power law
solution in this truncation scheme.  This feedback was first 
considered in Refs.~\cite{Maris:1995ns,Gusynin:1995bb,Maris:1996zg}.

\begin{figure}
\centerline{
\epsfig{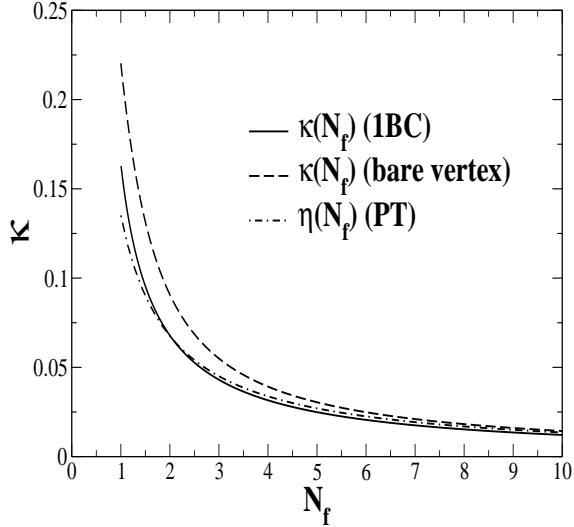}
}
\caption{\label{fig0} 
Here we display the anomalous dimension of the fermion vector
dressing function obtained from our infrared analysis, $\kappa(N_f)$,
compared with the conjecture from perturbation theory, $\eta(N_f)$.}
\end{figure}

An explicit numerical solution of the Eqs.~(\ref{ph-bc-kappa}) and
(\ref{A-IR-1BC}) is shown in Fig.~\ref{fig0}.  For the sake of
comparison we also display the solution for the case of a bare
fermion-photon vertex, which can be obtained from a similar
analysis.  Both results are very well fitted by a series of powers of
$1/N_f$: 
\beqa 
 \kappa_{bare} &=& \frac{0.135}{N_f} + \frac{0.090}{N_f^2} + 
 O(1/N_f^3) \, ,
\label{kappa_bare} 
\\ 
 \kappa_{1BC} &=& \frac{0.115}{N_f} + \frac{0.044}{N_f^2} + 
 O(1/N_f^3)  \, ,
\label{kappa_1BC}
\eeqa 
which suggests a connection to the $1/N_f$-expansion. 
Comparing the first term of our result for the bare vertex with
Eq.~(\ref{A-1/N}) we find that the $1/N_f$-result is indeed the first
term in the build up of an anomalous dimension.  The additional
$1/N_f^2$-term in our fit indicates that also loop corrections to the
next order in a $1/N_f$-expansion sum up and contribute to the
anomalous dimension.  However, one should keep in mind that our
calculation is not a $1/N_f$ expansion of the DSEs.  It is therefore
not surprising that our result for the order $1/N_f^2$ contribution to
the anomalous dimension deviates from that obtained in a $1/N_f$
expansion \cite{Gracey:1993iu}
\beqa
 \kappa_{1/N_f} &=& \frac{4}{3\pi^2 N_f}
    - \frac{8(32-3\pi^2)}{9\pi^4 N_f^2}
    +  O(1/N_f^3)
\\
    &\approx& \frac{0.135}{N_f} - \frac{0.022}{N_f^2}
    +  O(1/N_f^3) \,.
\eeqa
Furthermore, the vertex dressing of the $1BC$-vertex modifies
$\kappa(N_f)$ to quite some extent.

The infrared powers presented in Fig.~\ref{fig0} are not the only
solutions of the Eqs.~(\ref{ph-bc-kappa}) and (\ref{A-IR-1BC}). In the
range $-1/2 < \kappa <1$ we found a second solution which is
excellently fitted by
\beqa
 \kappa_{1BC}  &=& 0.5 -\frac{0.050}{N_f} - \frac{0.006}{N_f^2}
 - \frac{0.028}{N_f^3} \, .
\label{kappa_1BC_2}
\eeqa
However, contrary to the solution (\ref{kappa_1BC}), this solution
does depend very heavily on the projection method in the photon
equation, i.e. it depends on the parameter $\zeta$ introduced in
Eq.~(\ref{tensor}).  Furthermore, it does not connect to the
ultraviolet behaviour of the dressing functions, i.e. we do not find
numerical solutions of the DSEs interpolating between the infrared
behaviour, Eq.~(\ref{kappa_1BC_2}), and the ultraviolet asymptotic
behaviour given in subsection \ref{asymp-sec}.  We
therefore discard the solution (\ref{kappa_1BC_2}) in the following.

We have shown so far that in the symmetric phase the power law
(\ref{power-law}) leads to a self-consistent solution of the fermion
and photon DSEs, assuming that the vertex is dominated by its
$\gamma_\mu$-part.  The crucial question is, of course, whether the
power law survives when additional structure of the vertex is taken
into account.  That this is indeed the case for our choice of the
fermion-photon vertex can be shown by a simple dimensional analysis.
Plugging the power law into the CP-vertex, Eqs.~(\ref{BC}) and
(\ref{CP}), we find that with a vanishing B-function all terms in the
vertex depend on combinations of momenta which are of the same order
$p^{2\kappa}$ as the leading term.  Thus after integration the
CP-vertex will only change the coefficients of the right hand sides of
the fermion and photon DSEs, Eqs.~(\ref{ph-bc-ir}) and
(\ref{A-IR-1BC}), but not the general power law behaviour.  We thus
expect a modified function $\kappa_{CP}(N_f)$ as compared to
$\kappa_{1BC}(N_f)$ and $\kappa_{bare}(N_f)$.  That these
modifications are small will be confirmed by our numerical analysis in
Sec.~\ref{secIII} below.  Further modifications have to be expected
from including other transverse parts of the fermion-photon vertex and
it is by no means excluded that $\kappa$ finally becomes negative.  We
will further discuss this possibility later on in Sec.~\ref{LK-sec}.

\subsubsection{The chirally broken phase close to $N_f^{\mathrm crit}$}
Next we investigate the chirally broken phase close to the critical
value $N_f^{\mathrm crit}$ 
of the phase transition (assuming for now that there is a
chirally broken phase, and a critical value of $N_f$).  In this region
the dynamically generated fermion mass will be extremely small
compared to $\alpha$.  The momentum range $B(0) \ll p \ll \alpha$ will
dominate the integral on the right hand side of the DSE for the
B-function, Eq.~(\ref{B-eq}), and therefore the chirally symmetric
solutions for the photon polarisation and the dressing function $A$
\beqa
A(p^2) &=& c p^{2\kappa} \\
\Pi(p^2) &=& Z_2 \frac{\alpha}{c}\:w(\kappa) \:p^{-1-2\kappa} \, ,
\eeqa
determined in the last subsection, can be substituted in the integral
on the right hand side for all momenta.  Choosing again the case of the
1BC-vertex for simplicity we obtain
\beq
B(p^2) = Z_2 \frac{\alpha}{N_f \:\pi^3} \int d^3q 
 \frac{1}{k^2 + Z_2 \frac{\alpha w_{1BC}(\kappa)}{c} k^{1-2\kappa}} \,
 \frac{2 B(q)}{c^2 q^{2+4\kappa} + B(q)^2}\,
 \frac{c (p^{2\kappa} + q^{2\kappa})}{2} \,.
\eeq
The angular integrals are easily performed, and we obtain
\beq
 B(p^2) = \frac{2\:Z_2\:c\:\alpha}{p \: N_f\: \pi^2} \frac{1}{1+2\kappa} 
 \int_0^\infty dq \,q\,
 \frac{B(q)\,(p^{2\kappa} +q^{2\kappa})}{c^2 q^{2+4\kappa} + B^2(q)}\,
 \ln\left(\frac{Z_2 \frac{\alpha \:w_{1BC}(\kappa)}{c} + (p+q)^{1+2\kappa}}
{ Z_2\:\frac{\alpha w_{1BC}(\kappa)}{c}+ |p-q|^{1+2\kappa}}\right) \, .
\eeq
As we are interested in the momenta $p,q \ll \alpha$ the scale
$\alpha$ cuts off the integral and the logarithm can safely be
expanded.  Furthermore close to the phase transition the interesting
region is the one with $B^2(q) \ll c^2 q^{2+4\kappa}$, therefore the
equation can be linearised.  Thus we arrive at
\beq
B(p) =\frac{4}{w_{1BC}(\kappa)  \: N_f\:\pi^2} \int_0^\alpha dq
 \frac{B(q)(p^{2\kappa} +q^{2\kappa})}{q^{4\kappa}} 
 [\mbox{max}(p,q)]^{2\kappa-1} \,.
\label{bc_1} 
\eeq
The analysis of this type of equation is well known
\cite{Fukuda:zb,Appelquist:1988sr}.  The integral equation can be
solved directly by substituting the power law $B(p) \sim p^b$ 
and comparing coefficients on both sides.  On the other hand,
by converting the (nonlinear) integral equation into a differential
equation one obtains the boundary condition
\beq
\left[p\frac{dB(p)}{dp} + B(p) \right]_{p=\alpha} = 0 \,,
\label{bc_2} 
\eeq
which has to be satisfied by the power law solution.  Therefore a
nontrivial exponent $b$ has to be either complex or $b=-1$.

 From the integral equation, Eq.~(\ref{bc_1}), we obtain both the
chirally symmetric solution $B(p) \equiv 0$ and the nontrivial
solution
\beqa
\lefteqn{ b \; = \; -\frac{1}{2} + 2\kappa \label{Nc-eq} }
\nonumber \\  && \hbox{} 
\pm \frac{1}{2}\sqrt{(1-4\kappa+8\kappa^2) - 
 \frac{16 (1-2\kappa)}{\omega(\kappa)} - 
 4 \sqrt{\kappa^2(1-2\kappa)^2 -
 \frac{16 \,\kappa^2\,(1-2\kappa)}{\omega(\kappa)}
 + \frac{16 \, (1-2\kappa)^2}{(\omega(\kappa))^2}}}  \,,
\nonumber\\
\eeqa
where we have used the abbreviation $\omega(\kappa) := w_{1BC}(\kappa)
\pi^2 N_f$.  Above a critical value $N_f^{\mathrm crit}$ 
both solutions of the
exponent $b$ are in the interval $-1<b<0$ and not compatible with the
boundary condition Eq.~(\ref{bc_2}): The system is in the symmetric
phase.  Setting the discriminant of the outer root in Eq.~(\ref{Nc-eq})
equal to zero and using Eq.~(\ref{ph-bc-ir}) we find a critical number
of flavours of
\beq
 N_f^{{\mathrm crit}, \, 1BC} \approx 3.56 \, ,
\eeq
for the case of the 1BC-vertex.  For the sake of comparison we also
give the result in the bare vertex truncation
\beq
N_f^{{\mathrm crit,} \, bare} \approx 3.96 \, .
\eeq
Note that for $\kappa=0$ we have $A=1$, $\Pi(p^2) = \alpha/p$ and
$w_{1BC} \equiv 1$, and consequently recover the well-known result
from the $1/N_f$-expansion \cite{Appelquist:1988sr}
\beq
 b = -\frac{1}{2} \pm \frac{1}{2} \sqrt{1-\frac{32}{\pi^2 N_f}} \,,
\eeq
which gives  
\beq
N_f^{{\mathrm crit}, \,1/N_f} = 32/\pi^2 \approx 3.24 \,.
\eeq
(This limit also serves to determine the correct sign in front of the
inner root of Eq.~(\ref{Nc-eq}).)

Although the critical number of flavours is not too far away from the
old $1/N_f$ result there is a clear qualitative difference between
solutions from the full (coupled) set of DSEs and the one from an
$1/N_f$-expansion: The nonperturbative nature of the DSEs manifests
itself in the power law solution of the vector dressing function
$A(p^2)$, i.e. in $\kappa \not= 0$.  Such a behaviour can never be
obtained in a perturbative expansion and is in marked contrast to
the assumption $A=1$ employed in the leading order $1/N_f$-expansion.
Similar criticism has been raised in
Refs.~\cite{Pennington:1988jw,Atkinson:1989fp,Pennington:1990bx,Walsh:1990},
and is at the origin of a longstanding controversy.

Having determined the location of the phase transition we now
investigate the behaviour of the B-function for 
$N_f \rightarrow N_f^{\mathrm crit}$
from below.  Abbreviating
\beq
b = -\frac{1}{2} + 2\kappa \pm \frac{1}{2} f_{1BC}(N_f,\kappa) \,,
\eeq
the oscillatory solution of the linearised equation, Eq.~(\ref{bc_1}),
can be written in the form
\beq
B(p) = p^{-1/2+2\kappa} \sin\left[\frac{1}{2}\,f_{1BC}(N_f,\kappa)\, 
  (\ln(p/B(0))\,+\,\delta)\right]
\eeq
with a phase $\delta$ and the relevant scale for mass generation,
$B(0)$, in the logarithm.  Plugging this solution into the boundary
condition Eq.~(\ref{bc_2}) we take the limit 
$N_f \rightarrow N_f^{\mathrm crit}$ and
arrive at the condition
\beq
\frac{1}{2} f_{1BC}(N_f,\kappa) \, (\ln(p/B(0))\,+\,\delta) = 
 n\pi - \frac{1}{1+4\kappa} f_{1BC}(N_f,\kappa) \, .
\eeq
It can be shown \cite{Appelquist:1986fd,Appelquist:1988sr}
that the value $n=1$ gives the lowest vacuum energy.  Therefore
\beq
B(0) = \alpha \, e^{\left(\frac{2}{1+4\kappa}+\delta\right)} 
\exp\left[\frac{-2\pi}{f_{1BC}(N_f,\kappa)}\right] \,.
\eeq
We find an exponentially decreasing mass $M(0)=B(0)/A(0)$ at zero
momentum close to the chiral phase transition.  This is in agreement
with the numerical findings of Ref.~\cite{Maris:1996zg}, where the
same 1BC-vertex was used in both the fermion and the photon DSE.  A
similar expression describes the B-function in the bare vertex
truncation and in the $1/N_f$-expansion \cite{Appelquist:1986fd,
Appelquist:1988sr,Nash:1989xx}.  Our numerical study in
Sec.~\ref{secIII} will demonstrate that an exponential decrease of
$B(0)$ with a modified function $f(N_f,\kappa)$ will also emerge when
the CP-vertex is employed.  This type of exponential behaviour near
$N_f^{{\mathrm crit}}$, which is different from the usual first or
second order phase transition, is reminiscent of a conformal phase
transition \cite{Miransky:1996pd}.  Strictly speaking however, it is
not a conformal phase transition, because $\mbox{QED}_3$ is a
super-renormalisable theory with a dimensionful coupling constant, and
the conformal symmetry is broken in both the chirally symmetric and
the broken phase.

\subsection{\label{IR-nonlandau}
Infrared analysis in general linear covariant gauges}
Having determined the infrared behaviour of the photon polarisation
and the fermion dressing functions in Landau gauge we now turn to
general linear covariant gauges.  In the following we will investigate
whether our ansatz for the fermion-photon interaction is sophisticated
enough to generate gauge covariance of the photon polarisation and the
fermion propagator in the symmetric phase.  To this end we will follow
the strategy to first re-analyse the DSEs in general linear covariant
gauges and then compare our findings with the corresponding ones from
performing a LKFT of our Landau gauge solutions.  We will investigate
whether our truncation allows for power law solutions in general
linear covariant gauges and whether the LKFT is consistent with such a
scenario.

\subsubsection{The photon and fermion DSEs} 
In QED the photon polarisation is a gauge invariant object.  This is
evident from its LKFT, given below in Eq.~(\ref{ph-LK}).  Thus in
general a subtle interplay of the fermion propagators and the
fermion-photon interaction has to guarantee the invariance of the
photon polarisation in its DSE.  On the perturbative one-loop level,
it has been shown recently \cite{Bashir:2001vi}, that at least in the
symmetric phase there have to be terms in the transverse part of the
vertex that are explicitly dependent on the gauge parameter $\xi$.
These terms are not constraint by the WGTI, and are missing in the
vertex truncation investigated in this work.  For general gauges these
terms will be important in the photon DSE and we therefore cannot
expect the photon polarisation to be gauge invariant at our level of
truncation.

Assuming a $\xi$-dependent power law for the vector dressing function
of the fermions
\beq
A(p^2) = c(\xi) \:p^{2\kappa(\xi)} \,,
\label{power-law2}
\eeq
and employing the BC-vertex in the photon DSE we end up with the same
expression for $\Pi(p^2)$ as in Landau gauge,
\beqa
 \Pi(p^2)  &=:& Z_2 \:\frac{\alpha}{c(\xi)} \: \left(w_{1BC}(\kappa(\xi))
  + w_{2BC}(\kappa(\xi))\right) \: p^{-1-2\kappa(\xi)} \, . 
\label{ph2}
\eeqa
where $w_{1BC}$ has been given in Eq.~(\ref{ph-bc-kappa}) and
$w_{2BC}$ abbreviates contributions from the remaining term of the
BC-vertex.  The detailed form of $w_{2BC}$ can be calculated
analytically, but will not be given here as it is not important in the
following.  We note that on this level of truncation gauge invariance
of the photon polarisation in the infrared requires the exponent
$\kappa$ to be independent of the gauge parameter $\xi$.  It is
somewhat surprising that as a result, also the infrared behaviour of
the vector fermion dressing function $A$ is gauge independent, since
it is governed by the same exponent $\kappa$.  This may or may not be
an inconsistency in our truncation as this function $A$ is in general
a gauge dependent object.  At this stage of the investigation one may
hope that the dependence of $\kappa$ on $\xi$ is weak, leading to
qualitatively similar results at least in the vicinity of Landau
gauge.

Next we analyse the fermion DSE, Eq.~(\ref{DSE-exact}), with the
CP-vertex.  In the $\xi$-part of the equation the vertex is replaced
by the inverse fermion propagator according to the WGTI.  We then
obtain
\beqa
  A(p^2) &=&  Z_2 + 
 \frac{Z_2\: \alpha}{N_f\: \pi^3} \:\int d^3q \Bigg\{ \frac{1}{1+\Pi(k^2)}\, 
 \frac{1}{p^2\,k^2\,q^2 \, A(q^2)} \left(  -\frac{k^2}{2} + 
 \frac{(p^2-q^2)^2}{2k^2} \right) \times
\nonumber \\ && \hbox{\hspace*{4.cm} }
\left(\frac{A(p^2)+A(q^2)}{2} + 
  \frac{A(p^2)-A(q^2)}{p^2-q^2}\frac{p^2+q^2}{2}\right) 
\nonumber \\ && \hbox{}
 +  \frac{\xi}{p^2\,k^2\,q^2} \:\left[ \frac{A(p^2)}{A(q^2)}
      \left(\frac{p^2\:q^2}{2k^2}-\frac{p^4}{2k^2}+\frac{p^2}{2}\right) 
    + \left(\frac{p^2\: q^2}{2k^2}-\frac{q^4}{2k^2}+\frac{q^2}{2}\right) 
    \right] \Bigg\} \,.
\label{a-eq-xi}
\eeqa 
As has been noted in Ref.~\cite{Burden:gy} the CP-vertex leads to
gauge covariant DSEs in the quenched massless case. Indeed in this
limit, $\Pi(k^2)=0$, and we have the gauge covariant solution
\beq
 \frac{1}{A(p^2)} = 1 - \frac{e^2 \: \xi}{8 \: \pi \: p} 
  \arctan\left(\frac{8\pi p}{e^2 \xi}\right) \,,
\label{a-quen-sol}
\eeq
of Eq.~(\ref{a-eq-xi}).  In Landau gauge, this solution reduces to the
trivial solution $A(p^2) = 1$.  The question is: is there
a self-consistent power law solution for the unquenched case?

Substituting the power laws, Eqs.~(\ref{power-law2}) and (\ref{ph2}),
in Eq.~(\ref{a-eq-xi}), we have to distinguish several cases:
\begin{itemize}
\item[(a)]$\kappa > -1/2$:\\
  In this case the photon propagator contributes as
  \beq
  \frac{1}{1+\Pi_{1BC}(k^2)} 
   \approx \frac{c}{Z_2 \alpha w_{1BC}} \:k^{1+2\kappa} \,,
  \eeq
  and with substituted power laws the integration of the right hand side
  of Eq.~(\ref{a-eq-xi}) leads to
  \beqa
   c \: p^{2\kappa} &=&  Z_2 \: \frac{1}{N_f \pi^2} 
   \left\{ \frac{c\: p^{2\kappa}}{Z_2 \: (w_{1BC}+w_{2BC})} h(\kappa)
   + \frac{1}{p} \, \xi \,\frac{\Gamma(3/2-\kappa)\Gamma(1/2+\kappa)}
   {\Gamma(1+\kappa)\Gamma(1-\kappa)} \right\}  \,,
  \label{A-eq-IR-kap}	         
  \eeqa
  where $h(\kappa)$ abbreviates a combination of $\Gamma$-functions and
  hypergeometric functions, which need not to be specified here.  The
  longitudinal $1/p$-term dominates the right hand side for all gauges
  except Landau gauge and again we do not find a self-consistent power
  law to this level of truncation.  The only way to obtain such a
  solution with $\kappa>-1/2$ is the presence of a $\xi$-dependent
  transverse term in the vertex cancelling the $\xi$-dependent
  longitudinal piece\footnote{Based on a less rigorous analysis of the
  infrared a similar conjecture has already been made
  in \cite{Walsh:1990}.}.  Then the precise value of $\kappa$ is again
  determined by the coefficients of the remaining terms.
\item[(b)]$\kappa \le -1/2$, $\kappa\neq -1$:\\ 
  In this case the vacuum polarisation vanishes for $k \to 0$ and the
  photon propagator is proportional to $1/k^2$ in the infrared region.
  Integrating the right hand side of Eq.~(\ref{a-eq-xi}), the
  $\xi$-independent part vanishes, and we obtain
  \beqa
  c \: p^{2\kappa} &=& \frac{1}{p} \,  \xi \, 
   \frac{Z_2 \, \alpha}{N_f\: \pi} \, 
   \frac{\Gamma(3/2-\kappa)\Gamma(1/2+\kappa)}
   {\Gamma(1+\kappa)\Gamma(1-\kappa)} \, .
   \label{A-eq-IR-xi}	         
  \eeqa
  Note that this remaining $\xi$-dependent expression is
  exact (cf. the comments below Eq.~(\ref{DSE-exact})).  Matching the
  coefficients (naively) gives $\kappa=-1/2$, but due to the divergence
  of the coefficient on the right hand side this is not a solution.
  There might exist a potential solution with $\kappa = -1$, which we
  will consider separately below.  Here we conclude that there is no
  nontrivial self-consistent power law solution with 
  $-1 < \kappa \le -1/2$ within our truncation.  Any possible nontrivial 
  solutions in this range have to be generated by other transverse parts 
  of the vertex.  For example, a term proportional to $p^{2\kappa}$ 
  after integration with appropriate coefficients  would be more singular
  in the infrared than the $\xi$-piece and lead to a self-consistent
  nontrivial solution. 
\item[(c)] $\kappa = -1$\\ 
  As in case (b), the vacuum polarisation vanishes for $k \to 0$ and
  thus Eq.~(\ref{a-eq-xi}) becomes effectively quenched. In this case
  there is at least one solution, namely Eq.~(\ref{a-quen-sol}).  
  Certainly, this trivial solution will not survive when further
  parts of the transverse vertex will be taken into account, but it 
  serves well in the following to illustrate an important point: 
  interestingly, the infrared behaviour of this solution is
  \beq
    A(p^2) = 3 \left(\frac{e^2 \xi}{8\pi}\right)^2 \frac{1}{p^2} + O(p^4) \,,
  \eeq
  for $p \ll e^2 \xi /(8\pi)$ and thus $\kappa=-1$.  Nevertheless the 
  corresponding pure power law is not a self-consistent solution, as 
  can be seen from Eq.~(\ref{A-eq-IR-xi}): the right hand side vanishes 
  for $\kappa = -1$.  The reason for this behaviour is to
  be found in the appearance of the new scale $e^2 \xi /(8\pi)$ introduced
  by the gauge transformation from Landau gauge to general linear 
  covariant gauges.  For small values of the gauge parameter this new 
  scale divides the momentum range $0<p<\alpha$ into two regions and 
  in general we cannot expect the pure power law to describe the physics 
  in this whole region. 
\end{itemize}
In all cases considered so far we do not find a self-consistent power
law solution in non-Landau gauges.  Two possible reasons for this
behaviour have been identified.  Missing transverse parts of the
fermion-photon vertex could play an important role in these gauges.
Furthermore the appearance of the new scale $e^2 \xi /(8\pi)$ may
invalidate a simple power law ansatz in the infrared region.  In the
next subsection we will investigate, whether one can derive additional
information from the LKFT.

\subsubsection{ \label{LK-sec}
Landau--Khalatnikov--Fradkin transformation of the Landau gauge solution}
Assuming for the moment that the power law for the vector dressing
function is qualitatively correct in Landau gauge, we will now use the
LKFT to determine the corresponding solutions in other gauges.  The
transformation laws for the photon and fermion propagators are most
easily specified in coordinate space and we give the transformation
rules for the propagators in Euclidean space.  The photon propagator
$D_{\mu\nu}(x,\Delta)$ in general gauges can be obtained from its
transverse Landau gauge form $D_{\mu \nu}(x,0)$ by the transformation
law
\beq
D_{\mu \nu}(x,\Delta)  = 
 D_{\mu \nu}(x,0) + \partial_\mu \partial_\nu \Delta(x)  \,,
\label{ph-LK}
\eeq
with the arbitrary function $\Delta$.  The corresponding transformation
law for the fermion propagator is
\beq
S(x,\Delta) = S(x,0) \, e^{(\Delta(x)-\Delta(0))e^2}  \,.
\eeq
These transformation laws leave the DSE and the WGTI form invariant.  
In linear covariant gauges and in general dimension $d$ the function
$\Delta(x)$ is given by
\beq
\Delta(x) = -\xi \int \frac{d^dq}{(2\pi)^d} \frac{e^{-iq\cdot x}}{q^4} \,,
\eeq
which leads to the familiar form of the photon propagator in linear
covariant gauges
\beq
D_{\mu \nu}(p,\xi)  = 
  \left( \delta_{\mu \nu} - \frac{p_\mu p_\nu}{p^2}\right) 
  \frac{1}{p^2(1+\Pi(p^2))} + \xi \frac{p_\mu p_\nu}{p^4} \,,
\eeq
in momentum space with the gauge invariant photon polarisation
$\Pi(p^2)$.  Furthermore for $\mbox{QED}_3$ one obtains the
transformation law
\beq
 S(x,\xi) = S(x,0) \, e^{-x \xi e^2 /(8\pi)} \,,
\label{LK-S}
\eeq
for the fermion propagator in coordinate space.

In the symmetric phase of Landau gauge $\mbox{QED}_3$ we found the
power law solution
\beq
S(p,0) = \frac{i\pslash}{p^2} \frac{1}{c\:p^{2\kappa}} \,,
\eeq
which leads to the corresponding expression 
\beqa
S(x,0) &=& 
 \int \frac{d^3p}{(2\pi)^3} e^{-i p\cdot x} \frac{i\pslash}{p^2} 
            \frac{1}{c\:p^{2\kappa}} \,,
 \nonumber\\
       &=&  \frac{\Gamma(1-2\kappa) \sin(\kappa\pi)}{c\:4\pi^2} 
            \frac{1-2\kappa}{\kappa} \frac{\gamma_i x_i}{x^{3-2\kappa}} \,,
\eeqa
in coordinate space (all integration formulae used in this subsection
are given in Appendix \ref{integrals}).  Applying the transformation
(\ref{LK-S}) we transform the propagator to general gauges and perform
the inverse Fourier-transform
\beqa
 S(p,\xi) &=& 
  \frac{\Gamma(1-2\kappa) \sin(\kappa\pi)}{c\: 4\pi^2} 
 \frac{1-2\kappa}{\kappa}  \int d^3x e^{i p\cdot x}
 \frac{\gamma_i x_i}{x^{3-2\kappa}}e^{-x \xi e^2/(8 \pi)} 
\\
&=& \frac{i \pslash}{p^2} \frac{1}{c\:\cos(\kappa\pi)}
 \frac{2\kappa-1}{2\kappa}
\Bigg[\frac{\cos[2\kappa \arctan(p8\pi/(\xi e^2))]}
{\left[p^2+(\xi e^2/(8\pi))^2\right]^\kappa} 
\nonumber\\ && \hbox{\hspace*{5cm}}
-\frac{\sin[(2\kappa-1) \arctan(p8\pi/(\xi e^2))]}
{\left[p^2+(\xi e^2/(8\pi))^2\right]^{\kappa-1/2} 
\: p \:(2\kappa-1)}\Bigg] \,. \label{LK0}
\eeqa
Note that in performing the LKFT with the infrared power law alone we
have implicitly assumed that contributions from $p > \alpha= N_f
e^2/8$ have no significant influence on the behaviour of the
transformed propagator for $p \ll \alpha$.  Furthermore note that the
LKFT has introduced a new scale, $\xi e^2/(8\pi)$.  In order to be
consistent with the previous assumption we have to restrict the gauge
parameter to small values, i.e.  $0 \ll \xi e^2/(8\pi) \ll \alpha$.
We then obtain two momentum regions of interest where we can expand
our solution:
\beqa
A(p,\xi) &=&  \left\{ \begin{array}{l@{\quad \mbox{for} \quad}l}
c\:(p^2)^{-1} \:\cos(\kappa\pi) \frac{3}{1-4\kappa^2} 
 \left(\frac{\xi e^2}{8\pi}\right)^{2\kappa+2} &  p \ll  \frac{\xi e^2}{8\pi}  
\\
c \: (p^2)^{\kappa{ }} & p \gg  \frac{\xi e^2}{8\pi} \label{LK1}
\end{array}\right. 
\eeqa
As expected this expression smoothly connects to the Landau gauge
power law when $\xi \rightarrow 0$.  In all other linear covariant
gauges we obtain the Landau gauge power law for momenta $p \gg
\frac{\xi e^2}{8\pi}$.  Below this scale we find surprising agreement
with the LKFT of the trivial solution, Eq.~(\ref{A-eq-IR-xi}), i.e. a
power $\kappa = -1$.

\subsubsection{Self-consistent power law solutions in covariant gauges}
If we take the LKFT result based on the power law solution in Landau
gauge, Eq.~(\ref{LK1}), at face value and combine it with the
information we extracted from the infrared analysis of the coupled
fermion and photon DSEs, Eqs.~(\ref{ph2}) (which implicitly defines
$w_{INC}$ and $w_{2BC}$), (\ref{A-eq-IR-kap}), and (\ref{A-eq-IR-xi}),
two consistent scenarios of massless $\mbox{QED}_3$ with power law
behaviour in the infrared are possible:
\begin{itemize}
\item[I)] 
 Landau gauge behaves differently in the extreme infrared than any
 other linear covariant gauge.  In Landau gauge we have a power law
 with small positive or negative values of the exponent $\kappa$
 (dependent on the details of the vertex truncation).  In other gauges
 we essentially obtain the free solution with $\kappa=-1$ for $p \ll
 \frac{\xi e^2}{8}$ and the same power law as in the Landau gauge for
 momenta $\frac{\xi e^2}{8\pi} \ll p \ll \alpha$.  As the region $p
 \ll \frac{\xi e^2}{8\pi}$ can always be gauged away, all interesting
 infrared physics is already contained in the Landau gauge power law.
 In the DSE this scenario could be realised by a transverse term in
 the vertex which is proportional to the gauge parameter $\xi$ and
 leads to a term of the order $1/p^2$ in the infrared after
 integration of the fermion DSE.  A subtle cancellation of terms in
 the photon equation has to guarantee a gauge invariant photon
 polarisation.
\item[II)]
 Landau gauge behaves similar to other linear covariant gauges.  This
 entails that the solutions with small positive values for $\kappa$
 found in Sec.~\ref{ir-landau} are artifacts of the truncation scheme,
 and the true solution is a power law in the infrared with an exponent
 $\kappa=-1$ for all values of the gauge parameter $\xi$ including
 Landau gauge.  Then the gauge dependent scale $\frac{\xi e^2}{8\pi}$
 does not distinguish two momentum regions with different behaviour
 of the propagator.  In the DSE this scenario requires a transverse term 
 in the vertex which does {\em not} explicitly contain the gauge 
 parameter $\xi$ and leads to a term of the order $1/p^2$ in the infrared 
 after integration.  Then the dressing function $A$ and subsequently the
 photon polarisation would be gauge invariant in the infrared. The
 fermion and photon DSEs would effectively decouple for momenta
 $p \ll \alpha$. 
\end{itemize}
Both possibilities are consistent with our findings from the LKFT and
only a detailed analysis of the transverse parts of the fermion-photon
vertex can decide which one is realised in $\mbox{QED}_3$.

\section{\label{secIII}
Numerical results} 

In the previous section we performed in some detail an analytical
determination of the infrared behaviour of $\mbox{QED}_3$ close to and
above the phase transition assuming a power law behaviour of the
fermion vector dressing function.  Here we present our numerical
solutions of the unquenched system of DSEs in both the massive and the
massless phase, deferring the presentation of our results in quenched
$\mbox{QED}_3$ to appendix \ref{num-quench}.

\subsection{Unquenched results and phase transition in Landau gauge}

\begin{table}[b]
\begin{tabular}{|c||c|c|c|c|c|c|c|c|}
 & \multicolumn{8}{c|}{$-\langle \bar{\Psi}\Psi\rangle/e^4$}\\ \hline
 $N_f$        & 1.0 & 2.0 & 2.8 & 3.0 & 3.1 & 3.3 & 3.4 & 3.5 \\\hline
 1BC-vertex   & $7 \cdot 10^{-4}$    & $3.5 \cdot 10^{-5}$  
	      & $2.5 \cdot 10^{-8}$  & $3.9 \cdot 10^{-10}$ 
              & $1.5 \cdot 10^{-11}$ & & &       \\        
 CP-vertex    & $1.2 \cdot 10^{-3}$  & $1.3 \cdot 10^{-4}$
	      & $1.7 \cdot 10^{-6}$  & $2.6 \cdot 10^{-7}$ 
              & $8.9 \cdot 10^{-8}$  & $5.0 \cdot 10^{-9}$  
              & $7.4 \cdot 10^{-10}$ & $6.6 \cdot 10^{-11}$
\end{tabular}
\caption{\label{chirtab}
The chiral condensate (calculated via Eq.~(\ref{trace-condensate}))
obtained in the 1BC-vertex model and employing the CP-vertex in the
fermion-DSE and the BC-vertex in the photon-DSE.}
\end{table} 
\begin{figure}[t]
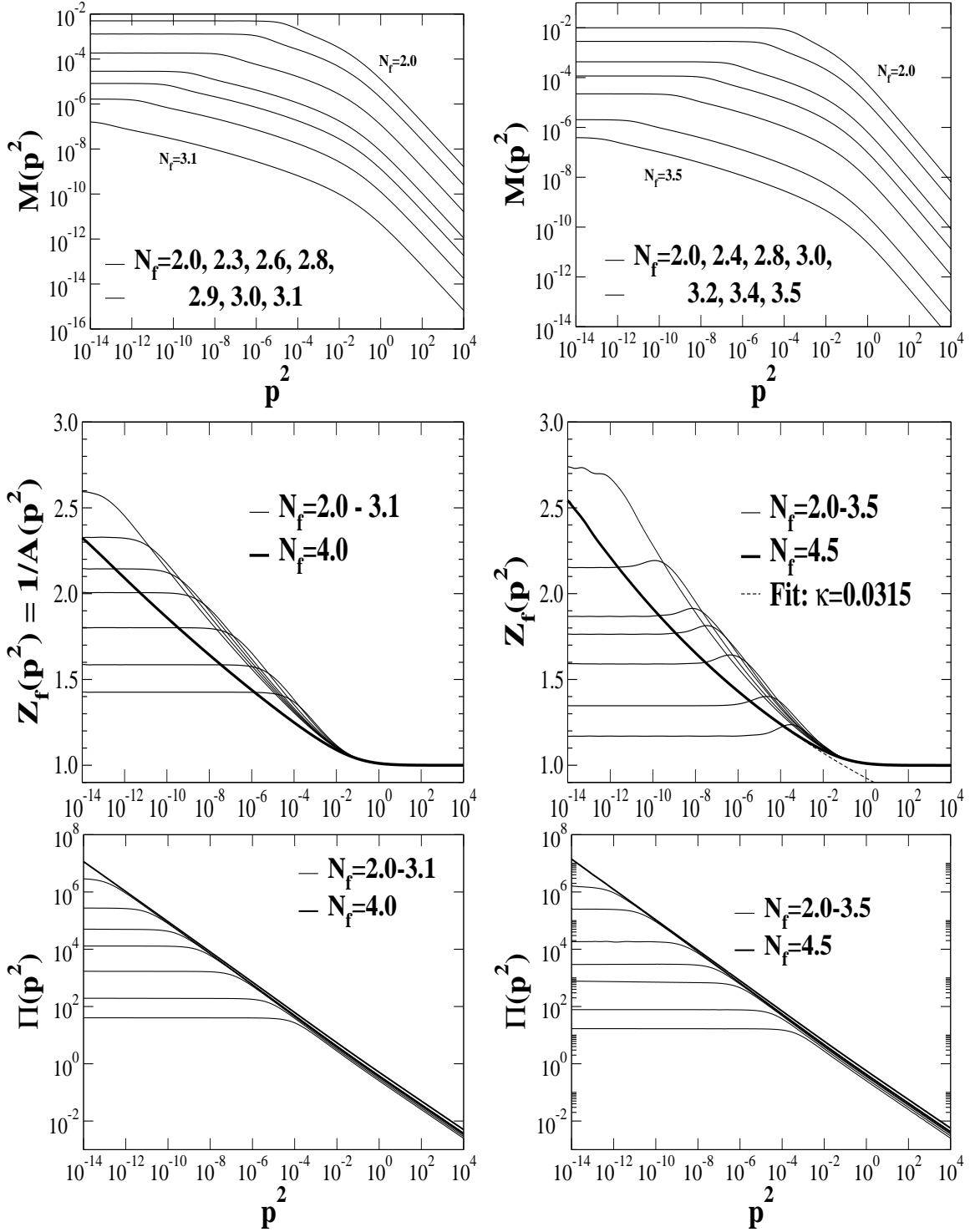

\parbox{8cm}{
\epsfig{file=./qed3-8.eps,width=7.5cm,height=6.5cm}
\epsfig{file=./qed3-9.eps,width=7.5cm,height=6.5cm}
\epsfig{file=./qed3-10.eps,width=7.5cm,height=6.5cm}
}\parbox{7.5cm}{
\epsfig{file=./qed3-8b.eps,width=7.5cm,height=6.5cm}
\epsfig{file=./qed3-9b.eps,width=7.5cm,height=6.5cm}
\epsfig{file=./qed3-10b.eps,width=7.5cm,height=6.5cm}
}
\caption{\label{fig4} 
Shown are the variation of the mass function, the wave function
renormalisation and the polarisation with the number of flavours in
Landau gauge, $\xi=0$. On the left hand side we employed the first
term of the BC-vertex (1BC) in both the photon and the fermion DSEs.
On the right hand side we display the same functions calculated with
the CP-vertex in the fermion DSE and the BC-vertex in the photon
DSE. The scale is set by choosing $e^2=1$.}
\end{figure}
\begin{figure}
\vspace{1cm}
\centerline{
\epsfig{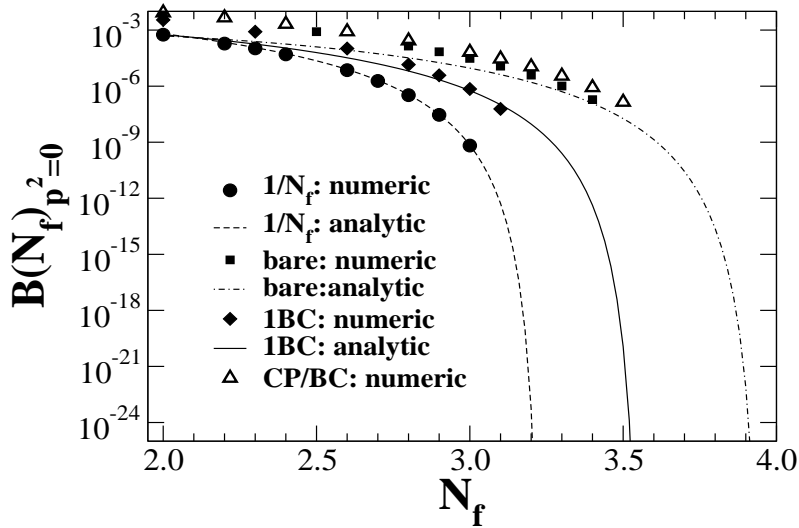}
}
\caption{\label{fig2a} 
The phase transition: Analytical vs. numerical results for the scalar
fermion dressing function, $B(p^2=0)$, as function of $N_f$.  Shown are
results for four different truncation schemes: The $1/N_f$-expansion
of Ref.~\cite{Appelquist:1988sr}, and three different vertex
truncations of the coupled photon and fermion DSEs.  The scale is set
by choosing $e^2=1$.}
\end{figure}

Our results for the fermion mass function $M(p^2)$, the wave function
renormalisation $Z_f(p^2)$ and the photon polarisation $\Pi(p^2)$ in
unquenched Landau gauge are shown in Fig.~\ref{fig4}.  On the left
panel we display results obtained with the first term of the BC-vertex
(1BC) only.  On the right panel we give the results obtained with the
CP-vertex in the fermion DSE and the BC construction in the photon
DSE.  All results in this section are obtained with the
Brown--Pennington projection $\zeta=3$ in the photon equation, other
choices lead only to minor modifications.  As expected, in the
ultraviolet all curves follow their respective asymptotic limits,
given in Eqs.~(\ref{UV_B}) and (\ref{UV_PH}).  In the infrared we find
finite, nonzero dressing functions throughout the dynamically broken
phase.  Furthermore, we can clearly see two distinct mass scales: The
fermion dressing functions have a kink near $p = e^2$ and a second
kink at $p \sim M(0)$.  Close to the phase transition, these scales
are several orders of magnitude apart, which makes lattice simulations
extremely difficult.

Above $N_f^{\mathrm crit}$ the functions turn into power laws thus
justifying the basic assumption in our IR-analysis of
Sec.~\ref{IR-sec}.  The values of the exponents $\kappa$ determined in
the analytical calculation are reproduced on the $10\%
$ level by the
numerical results.  In the CP/BC-vertex case we find $\kappa=0.0315$
at $N_f=4.5$, which is much closer to the value obtained with a bare
vertex, $\kappa_{bare}=0.0343$, than to the corresponding value
$\kappa_{1BC}=0.0278$ obtained with the 1BC-vertex piece.
Furthermore, notice that for these unquenched calculations, $Z_f(p^2)
\ge 1$, at least for $N_f \ge 2$, in contrast to our findings in the
quenched case (cf appendix \ref{num-quench}).

The dynamical mass generation close to the phase transition is studied
in Fig.~\ref{fig2a}.  Shown are results in four different truncation
schemes: The leading order $1/N_f$-expansion of
Ref.~\cite{Appelquist:1988sr} employs the perturbative expression,
Eq.~(\ref{UV_PH}), for the photon polarisation and chooses
$A(p^2)\equiv 1$ together with a bare fermion photon vertex.  This is
compared to our results from the fully unquenched system of DSEs with
three different truncations for the vertex: bare, 1BC and the CP/BC
combination.  As can be seen from the figure, the numerical results
follow nicely the corresponding analytical results from
Sec.~\ref{ir-landau}, and are in qualitative agreement with the
findings of Ref.~\cite{Maris:1996zg}.  Our most sophisticated vertex
truncation, the CP/BC-combination, tends toward a critical value of
the number of flavours of $N_f^{\mathrm crit} \approx 4$.

Finally we list our results for the chiral condensate in two different
truncation schemes for a range of values of $N_f$ in
Table~\ref{chirtab}.  For the CP-vertex, there is no discrepancy
between the condensate as obtained from the trace of the fermion
propagator
\beq
 \langle \bar{\Psi} \Psi \rangle = -\mbox{Tr}[S(0)] = 
  -4 \int \frac{d^3q}{(2 \pi)^3} 
 \frac{B(q^2)}{q^2 A^2(q^2) + B^2(q^2)} \,,
\label{trace-condensate}
\eeq
and that extracted from the asymptotic behaviour, Eq.~(\ref{UV_B});
for the 1BC-vertex there is about 5\% to 10\% difference between the
two methods.  Listed are the condensates calculated from the trace of
the fermion propagator.  In accordance with the simple estimate given
in Ref.~\cite{Appelquist:2004ib}, we find small condensates well below
the phase transition.  For $N_f=1$ and $N_f=2$ recent lattice
simulations provide upper bounds of the $O(10^{-3}) e^4$ and $O(10^{-4})
e^4$, respectively \cite{Hands:2002dv, Hands:2004bh}.  These bounds
are certainly consistent with our values.  Thus the combined evidence
of the DSEs and the lattice Monte-Carlo simulations indicate the
presence of dynamical chiral symmetry breaking at $N_f=1$ and $N_f=2$.

\subsection{Unquenched results in general linear covariant gauges}
%
\begin{figure}
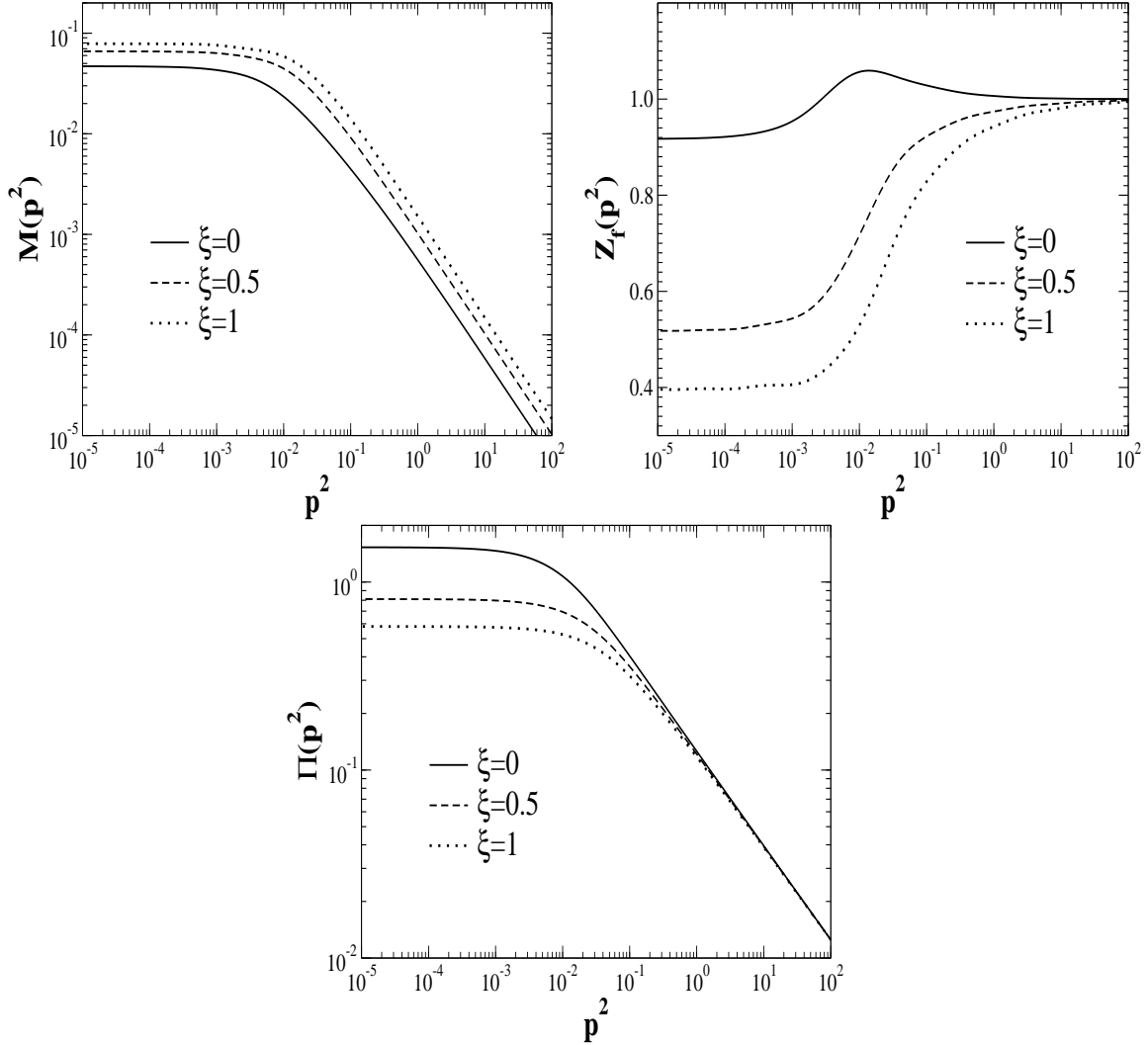

\parbox{8cm}{
\epsfig{file=./qed3-5.eps,width=7.5cm,height=7cm}
}\parbox{7.5cm}
{\epsfig{file=./qed3-6.eps,width=7.5cm,height=7cm}}
\centerline{
\epsfig{file=./qed3-7.eps,width=7.5cm,height=7cm}
}
\caption{\label{fig3}
Shown are our results for the dressing functions for three different 
values of the gauge parameter $\xi$ and $N_f = 1$.  
The scale is set by choosing $e^2=1$.}
\end{figure}
\begin{figure}
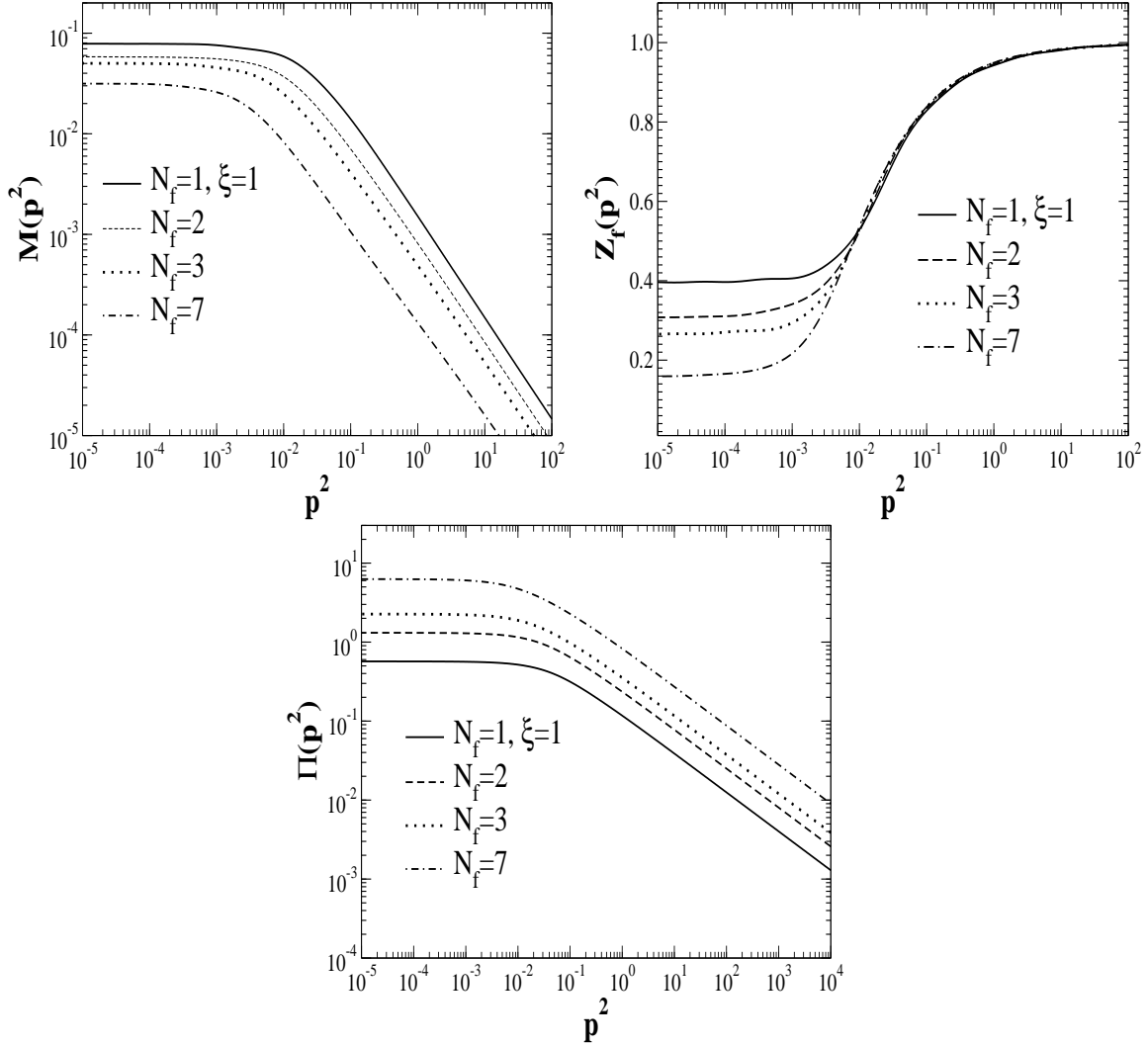

\parbox{8cm}{
\epsfig{file=./qed3-11.eps,width=7.5cm,height=7cm}
}\parbox{7.5cm}{\epsfig{file=./qed3-12.eps,width=7.5cm,height=7cm}
}
\centerline{
\epsfig{file=./qed3-13.eps,width=7.5cm,height=7cm}
}
\caption{\label{fig5} 
Shown are the variation of the mass function, the wave function
renormalisation and the polarisation with the number of flavours in
Feynman gauge, $\xi=1$.  The scale is set by choosing $e^2=1$.}
\end{figure}
\begin{table}
\begin{tabular}{|c||c|c|c|c|c|c|c|}
gauge parameter 
& \multicolumn{7}{c|}{$-\langle \bar{\Psi}\Psi\rangle/(10^{-5} e^4)$}\\ \hline
& $N_f=0$ & $N_f=1$ & $N_f=2$ & $N_f=3$ & $N_f=4$ & $N_f=5$ & $N_f=6$ \\ \hline
$\xi=0$  & 333  & 121     & 13      & 0.026   & ?       & 0       & 0  \\ 
$\xi=0.5$& 340  & 165     & 79      & 39      & 23      & 15      & 11  \\
$\xi=1$  & 351  & 202     & 108     & 74      & 55      & 37      & 29   \\
$\xi=2$  & 356  & 259     & 189     & 143     & 107     & 92      & 77 
\end{tabular}
\caption{\label{qtable3} 
The chiral condensate (calculated via Eq.~(\ref{trace-condensate}))
obtained in the CP/BC-vertex truncation in
different gauges.}
\end{table} 

Here we present numerical solutions for unquenched $\mbox{QED}_3$ in
linear covariant gauges in the CP/BC-vertex truncation scheme.
According to our previous discussion in Sec.~\ref{IR-nonlandau} and
the numerical results in the quenched approximation in Appendix
\ref{num-quench}, we expect artifacts from violating gauge invariance.
The results presented in Fig.~\ref{fig3} for the case of $N_f=1$
flavours indicate that this is indeed the case.  The photon
polarisation functions clearly depend on the gauge parameter $\xi$.
On a quantitative level this can also be seen from the values of the
chiral condensate displayed in Table~\ref{qtable3}.  Induced by the
feedback of the photon propagator on the fermion, the variation of the
condensate with the gauge parameter is much larger for the unquenched
than for the quenched theory.  Only the Landau gauge results are
consistent with the bounds from lattice simulations.

As expected from the infrared analysis of Sec.~\ref{IR-nonlandau}
there is no chirally symmetric self-consistent power law solution for
$\xi \ne 0$.  In fact we did not find any self-consistent solution in
the symmetric phase for $\xi \ne 0$, and consequently the system stays
in the chirally broken phase at least for values of $N_f$ as large as
$7$, which is the largest $N_f$ we have investigated.  For larger
values of $N_f$ the numerical analysis becomes increasingly tedious.
The corresponding dressing functions in Feynman gauge are presented in
Fig.~\ref{fig5}.  An interesting difference with the Landau gauge
solutions is that now $0 < Z_f(p^2) \le 1$ on the entire momentum
range, in contrast to the unquenched Landau gauge solutions, for which
$Z_f(p^2) \ge 1$ on part ($N_f=1$) or all ($N_f \ge 2$) of the 
momentum range.  Also note that now there appears to be only 
one scale at which the generated fermion mass function function
$M(p^2)$ has a kink.

\section{\label{summary}
Summary and Conclusions}
In this work we have investigated the chiral phase transition of
$\mbox{QED}_3$ in the Green's functions approach.  Employing different
ans\"atze for the fermion-photon vertex we have solved the
coupled set of Dyson--Schwinger equations for the fermion and photon
propagators.  No other approximations have been made in our numerical
calculations.  In addition, the infrared behaviour of the propagators
close to the phase transition and in the symmetric phase has been
investigated employing methods that have been successfully used
previously in four-dimensional QCD \cite{vonSmekal:1997is,
Atkinson:1998tu,Lerche:2002ep,Fischer:2002hn,Fischer:2003rp} and QED
\cite{Fukuda:zb}.  Special care has been taken to preserve gauge
invariance as much as possible.  The chosen vertex ansatz satisfies
the Ward--Green--Takahashi identity.  Furthermore it has the correct
properties under Landau--Khalatnikov--Fradkin transformations in the
special case of massless quenched QED.  Nevertheless, our results
indicate that this is not enough: the resulting vacuum polarisation is
not gauge invariant, nor is the chiral condensate.  Clearly, further
structure in the transverse part of the vertex is needed to properly
ensure gauge covariance.

In Landau gauge we find a self-consistent power law solution for the
photon polarisation and the vector fermion dressing function in the
infrared region in the symmetric phase.  With a bare fermion-photon
vertex the anomalous dimension $\kappa$ is directly related to the
coefficients of a well known result from the $1/N_f$-expansion.  We
thus confirmed a longstanding conjecture from the renormalisation
group \cite{Appelquist:1981sf,Atkinson:1989fp,Pennington:1990bx}.  We
would like to emphasise, however, that such a power law solution is
genuinely nonperturbative in nature and cannot be obtained to any
finite order in perturbation theory or the $1/N_f$-expansion.  The
dependence of $\kappa$ on the number of flavours is modified by
nonperturbative contributions in the vertex dressing. We find small
positive values of $\kappa$ for all vertex dressings considered so
far, indicating a vanishing propagator function $A(p^2)$ for $p^2 \to
0$.  This is not what one would expect on physical grounds
\cite{Franz:2002qy}.  It remains to be seen whether further
contributions from the transverse part of the fermion-photon vertex
are capable to drive $\kappa$ to negative values.

In the chirally broken phase, the power law behaviour gets modified by
the (dynamically generated) fermion mass, which effectively acts as an
infrared cutoff.  We have determined the dependence of the chiral
condensate and of the scalar fermion dressing function $B(p^2=0)$ on
the number of flavours and found an exponential decrease close to the
phase transition.  If this behaviour turns out to be correct in the
full theory, one can hardly hope to be able to determine the critical
number of flavours from lattice Monte-Carlo simulations.  A sign
pointing in this direction is found for small $N_f$: The chiral
condensate is very small compared to the natural mass scale $e^2$.
Furthermore it agrees with the values recently determined on the
lattice \cite{Hands:2002dv,Hands:2004bh}.  The qualitative behaviour
of the B-function and the condensate does not depend on our choice for
the vertex ansatz, though there are quantitative differences, in
particular near the critical number of flavours, $N_f^{\mathrm crit}$.
The results with our most sophisticated vertex suggest a critical
number of flavours $N_f^{\mathrm crit} \approx 4$.

In other linear covariant gauges we find a completely different
picture.  No indications for a phase transition are seen in our
numerical analysis.  The value of the condensate becomes heavily
dependent on the gauge parameter, and for $\xi \neq 0$ our results
exceed the bounds set by lattice simulations.  It appears as if no
self-consistent power law solutions exist in the symmetric phase.  A
possible explanation for this fact can be found with the help of the
Landau--Kalatnikov--Fradkin transformation laws.  We find that, given
a power law solution in Landau gauge with exponent $\kappa$, the gauge
transformed propagator has the same anomalous dimension only for
momenta $ \frac{\xi e^2}{8\pi} \ll p \ll \alpha$, whereas below this
scale it has an anomalous dimension $\kappa = -1$.  Such a solution
cannot be found from the Dyson--Schwinger equations with our vertex
truncation.  We thus conclude that further transverse structure in the
vertex is mandatory to obtain gauge covariant solutions for the
propagators of $\mbox{QED}_3$.  These extra terms could allow for
three possible scenarios in the symmetric phase of $\mbox{QED}_3$: (a)
terms explicitly proportional to the gauge parameter $\xi$ could lead
to solutions in accordance with the Landau--Khalatnikov--Fradkin
transformation of the Landau gauge solutions; or (b) they could
invalidate all Landau gauge solutions found so far; or (c) they could
allow for solutions in the symmetric phase that are not power laws in
the infrared region.  Although we do believe that the scenario (a) is
most likely to be realised we cannot exclude the other possibilities
so far.  A thorough investigation of all existing proposals for the
transverse structure of the fermion-photon vertex in this respect
seems desirable and will be carried out in future work.

\section*{Acknowledgement}
We are grateful to H.\ Gies, J.\ Pawlowski and M.\ Pennington for
useful discussions and a critical reading of the manuscript.  We thank
B.\ Gr\"uter, S.\ Ahlig, and F.\ Kren for helpful communications on
numerical issues, and S.\ Hands, I.\ Herbut, N.\ Mavromatos and Z.\
Tesanovic for valuable discussions.  This work has been supported by a
grant from the Ministry of Science, Research and the Arts of
Baden-W\"urttemberg (Az: 24-7532.23-19-18/1 and 24-7532.23-19-18/2)
and by the Deutsche Forschungsgemeinschaft under contract Fi 970/2-1.

\begin{appendix}
\section{\label{app1}
Regularisation of the photon DSE}
The aim here is to provide a formulation for the photon equation that
can be used not only in combination with the Brown--Pennington
projector, $\zeta=3$ (cf. Eq.~(\ref{tensor})), but with general values
of $\zeta$.  The dependence of the Landau gauge vacuum polarisation on
$\zeta$ then provides a measure of the violation of transversality in
the photon equation.

To see the problems arising with $\zeta \not= 3$ we analyse the
ultraviolet behaviour of the photon DSE, Eq.~(\ref{quarkloop}). The
kernels $W_i$ are then given by
\beqa
W_1(p^2,q^2,k^2) &=&\frac{\zeta k^4}{p^4} + k^2\left(\frac{1-\zeta}{p^2} - \frac{2\zeta q^2}{p^4}\right)
              -1+\frac{(1-\zeta)q^2}{p^2}  + \frac{\zeta q^4}{p^4} \,, \\
W_2(p^2,q^2,k^2) &=& \frac{2 (3-\zeta)}{p^2} \,, \\	       
W_3(p^2,q^2,k^2) &=& \frac{\zeta k^6}{p^4} - k^4\left(\frac{1+\zeta}{p^2} + \frac{\zeta q^2}{p^4}\right)
              +k^2 \left(1 + \frac{2(\zeta-3)q^2}{p^2}-\frac{\zeta q^4}{p^4}  \right) \nonumber\\
	    &&+ q^2 - \frac{(\zeta+1)q^4}{p^2} + \frac{\zeta q^6}{p^4} \,, \\
W_4(p^2,q^2,k^2) &=& \frac{-2 \zeta k^4}{p^4} + k^2 \left(\frac{4}{p^2}+\frac{4 \zeta q^2}{p^4}  \right) 
              -2 + \frac{4q^2}{p^2} - \frac{2 \zeta q^4}{p^4}  \,,\\ 
W_5(p^2,q^2,k^2) &=&  \frac{\zeta k^4}{p^4} -k^2\left( \frac{1+\zeta}{p^2}+\frac{2\zeta q^2}{p^4}\right)
              + 1 + \frac{(\zeta-3) q^2}{p^2} + \frac{\zeta q^4}{p^4}  \,,
\eeqa
\beqa
W_6(p^2,q^2,k^2) &=& \frac{\zeta k^4}{p^4} - k^2 \left(\frac{3-\zeta}{p^2}+\frac{2\zeta q^2}{p^4} \right)
              + 1  + \frac{(-\zeta-1) q^2}{p^2}+ \frac{\zeta q^4}{p^4} \,.
\eeqa
Let us first concentrate on the term proportional to the kernel $W_1$.
In the ultraviolet we can use the angular approximation
$A(q^2)=A(k^2)=1$, furthermore the momenta $q^2$ and $k^2$ are larger
than all fermion masses.  One angular integral yields a trivial factor
$2\pi$ and the integral is dominated by the part $q^2 > p^2$
\beqa
 \Pi^{UV1}(p^2) &=&-\frac{g^2 N_f}{(2 \pi)^2} \int_{p}^\infty dq q^2 
 \int_0^{\pi} d\theta \sin(\theta) \: \frac{1}{q^2 k^2} 
 \times \nonumber\\
&& \left[\frac{\zeta k^4}{p^4} + k^2\left(\frac{1-\zeta}{p^2} - \frac{2\zeta q^2}{p^4}\right)
              -1+\frac{(1-\zeta)q^2}{p^2}  + \frac{\zeta q^4}{p^4}\right]. 
\eeqa
Now we perform the angular integrals according to Eqs.~(\ref{1})
through (\ref{3}) and expand the resulting logarithm for momenta $q^2
\gg p^2$.  To leading order we obtain
\beqa
 \Pi^{UV1}(p^2)
&=& -\frac{g^2 N_f}{(2 \pi)^2} \int_{p}^\infty dq  
 \left\{ 
 \frac{2(3-\zeta)}{3 p^2} + \frac{(-10-2\zeta)}{15q^2} + O(p^2/q^4) 
 \right\} \,,
\eeqa
which is convergent iff $\zeta=3$ but contains a linear divergence for
all other values.  Treating all other terms in the fermion loop in the
same way we arrive at the expression
\beqa
 \Pi^{UV}(p^2) &=& -\frac{g^2 N_f}{(2 \pi)^2} \int_{p}^\infty   
dq \Bigg\{ 
 \frac{2(3-\zeta)}{3 p^2} + \frac{(-10-2\zeta)}{15q^2} + \cdots 
+ B(q^2)^2\left(\frac{2(3-\zeta)}{p^2q^2}
+\frac{2(3-\zeta)}{3q^4}\right)\nonumber \\
& +& \frac{A^\prime(q^2)}{2}\left[-\frac{4q^2(3-\zeta)}{3p^2} + \frac{5-\zeta}{15} + \cdots
+B(q^2)^2\left(\frac{4(3-\zeta)}{3p^2} -\frac{5+3\zeta}{15q^2} + \cdots
\right)\right]\nonumber\\
&+& B^\prime(q^2)A(q^2)B(q^2) \left[-\frac{8(3-\zeta)}{3p^2} +\frac{10+6\zeta}{15q^2} + \cdots
\right] \Bigg\} \; .
\eeqa
which contains linear divergences proportional to $(3-\zeta)$ at
various places. In order to eliminate these terms we subtract
appropriate expressions from the kernels $W_i$ given in
Eqs.~(\ref{w1}) through (\ref{w6}).  This results in the modified
kernels
\beqa
\widetilde{W}_1(p^2,q^2,k^2) &=& W_1(p^2,q^2,k^2) - \frac{2k^2(3-\zeta)}{3p^2} \,, \\
\widetilde{W}_2(p^2,q^2,k^2) &=& 0 \,, \\
\widetilde{W}_3(p^2,q^2,k^2) &=& W_3(p^2,q^2,k^2) + \frac{8q^2k^2(3-\zeta)}{3p^2} \,, \\
\widetilde{W}_4(p^2,q^2,k^2) &=& W_4(p^2,q^2,k^2) - \frac{8k^2(3-\zeta)}{3p^2} \,, \\
\widetilde{W}_5(p^2,q^2,k^2) &=& W_5(p^2,q^2,k^2) + \frac{4k^2(3-\zeta)}{3p^2} \,, \\
\widetilde{W}_6(p^2,q^2,k^2) &=& W_6(p^2,q^2,k^2) + \frac{4k^2(3-\zeta)}{3p^2} \,.
\eeqa
Based on our analytical infrared calculus as well as on our numerical
calculations we investigated the dependence of the solutions on the
projection parameter $\zeta$ and found very small effects not
affecting any of our conclusions in the main body of this work.

\section{\label{integrals}
Angular and radial integrals} 
In $d=3$ dimensions the following angular integrals are needed for the
UV-analysis of the photon equation
\beqa
\int_0^\pi d\theta \frac{\sin(\theta)}{k^4} &=& \frac{2}{(q^2-p^2)^2} \,,
\label{1}\\
\int_0^\pi d\theta \frac{\sin(\theta)}{k^2} &=&  
       \frac{1}{pq}\ln\left(\frac{p+q}{|p-q|}\right) \,,
       \\
\int_0^\pi d\theta \sin(\theta) &=&  2 \,, \\
\int_0^\pi d\theta \sin(\theta) \: k^2 &=&  2(p^2+q^2) \,, \\
\int_0^\pi d\theta \sin(\theta) \: k^4 &=&  2p^4+2q^4+\frac{20}{3}p^2q^2 \,,
\label{3}
\eeqa
where $k^2=(q-p)^2 = p^2+q^2-2pq\cos(\theta)$.  For the IR-analysis of
the coupled system we need the following integrals
\beqa
\int d^dq \, \frac{1}{(q^2)^a (k^2)^b} &=&  
\pi^{d-d/2} (p^2)^{d/2-a-b} \frac{\Gamma(d/2-a) \Gamma(d/2-b) \Gamma(a+b-d/2)}
{\Gamma(a) \Gamma(b) \Gamma(d-a-b)} \,,
\label{IR-int}
\eeqa
and for the Fourier-transforms necessary for the LKFT we need
\beqa
\int_0^\pi d\theta \sin\theta \cos\theta \, e^{\pm i\:px\: \cos\theta} &=& 
\mp 2i \left(\frac{\cos(px)}{px} - \frac{\sin(px)}{(px)^2}\right) \,,
\\
\int_0^\infty dx\, x^b \,\sin(ax) &=& \frac{\Gamma(1+b)}{a^{1+b}} \sin\left[(1+b)\frac{\pi}{2}\right], 
\,\,\,\, 0<|b+1|<1 \,, \,\,\,\,[3.761(4)]
\\
\int_0^\infty dx \,x^b\, \cos(ax) &=& \frac{\Gamma(1+b)}{a^{1+b}} \cos\left[(1+b)\frac{\pi}{2}\right],
\,\,\,\, 0<(b+1)<1 \,, \,\,\,\,[3.761(9)]
\eeqa
\beqa
\int_0^\infty dx\, x^b\, \sin(ax) \, e^{-cx} &=& \frac{\Gamma(1+b)}{(a^2+c^2)^{(1+b)/2}} 
\sin\left[(1+b)\arctan\left(\frac{a}{c}\right)\right], \,\,\,\, b>-2, \: c>0 \,, \nonumber\\
&& \hspace*{7.3cm}[3.944(5)] \\
\int_0^\infty dx \,x^b\, \cos(ax) \, e^{-cx} &=& \frac{\Gamma(1+b)}{(a^2+c^2)^{(1+b)/2}} 
\cos\left[(1+b)\arctan\left(\frac{a}{c}\right)\right], \,\,\,\, b>-1, \: c>0 ,\,, \nonumber\\
&& \hspace*{7.3cm}[3.944(6)]
\eeqa
where the numbers in square brackets refer to the corresponding
equations in Ref.~\cite{Grad}.

\section{\label{num-quench}
Numerical results in quenched approximation} 

$\mbox{QED}_3$ in quenched approximation employing the BC-vertex in
the fermion and photon equation has been investigated in detail in
Refs.~\cite{Burden:mg, Burden:1991uh} (for recent work see
Ref.~\cite{Bashir:2004yt} and references therein).  In the chirally
broken phase, the gauge dependence of the photon polarisation as well
as the chiral condensate was found to be rather weak for the
condensate and uncomfortably large for the photon polarisation.  For
the condensate a discrepancy between the value extracted from the
asymptotic form of the scalar fermion self-energy, see
Eq.~(\ref{UV_B}), and the value obtained from the trace of the fermion
propagator, see Eq.~(\ref{trace-condensate}), has been found.  All we
have to add to these investigations is an answer to this last problem.
As can be seen from Table~\ref{qtable}, adding the CP term to the
BC-vertex in the fermion DSE removes this discrepancy and slightly
reduces the gauge dependence of the condensate.
%
\begin{table}[b]
\begin{tabular}{|c||c|c||c|c||c|c|}
$(-\langle \bar{\Psi}\Psi\rangle)/(10^{-3} e^4$)
           & \multicolumn{2}{c||}{$\xi=0$  } & 
	     \multicolumn{2}{c||}{$\xi=0.5$} & 
	     \multicolumn{2}{c|}{$\xi=1$  }  \\ \hline
             & asympt. & -Tr[S(0)] 
	     & asympt. & -Tr[S(0)] 
	     & asympt. & -Tr[S(0)] \\ \hline
 BC-vertex &  3.34   &  3.24     &  3.54   &   3.25    & 3.67    &  3.26 \\
 CP-vertex &  3.29   &  3.29     &  3.34   &   3.35    & 3.46    &  3.46 
\end{tabular}
\caption{\label{qtable} 
The chiral condensate extracted from the asymptotics of the fermion
mass function, see Eq.~(\ref{UV_B}), and obtained by taking the trace
of the propagator, for different values of the gauge parameter $\xi$,
all in the quenched ($N_f=0$) approximation.  The units are $10^{-3}
e^4$.}
\end{table} 
\begin{figure}[t]
\parbox{8cm}{\epsfig{file=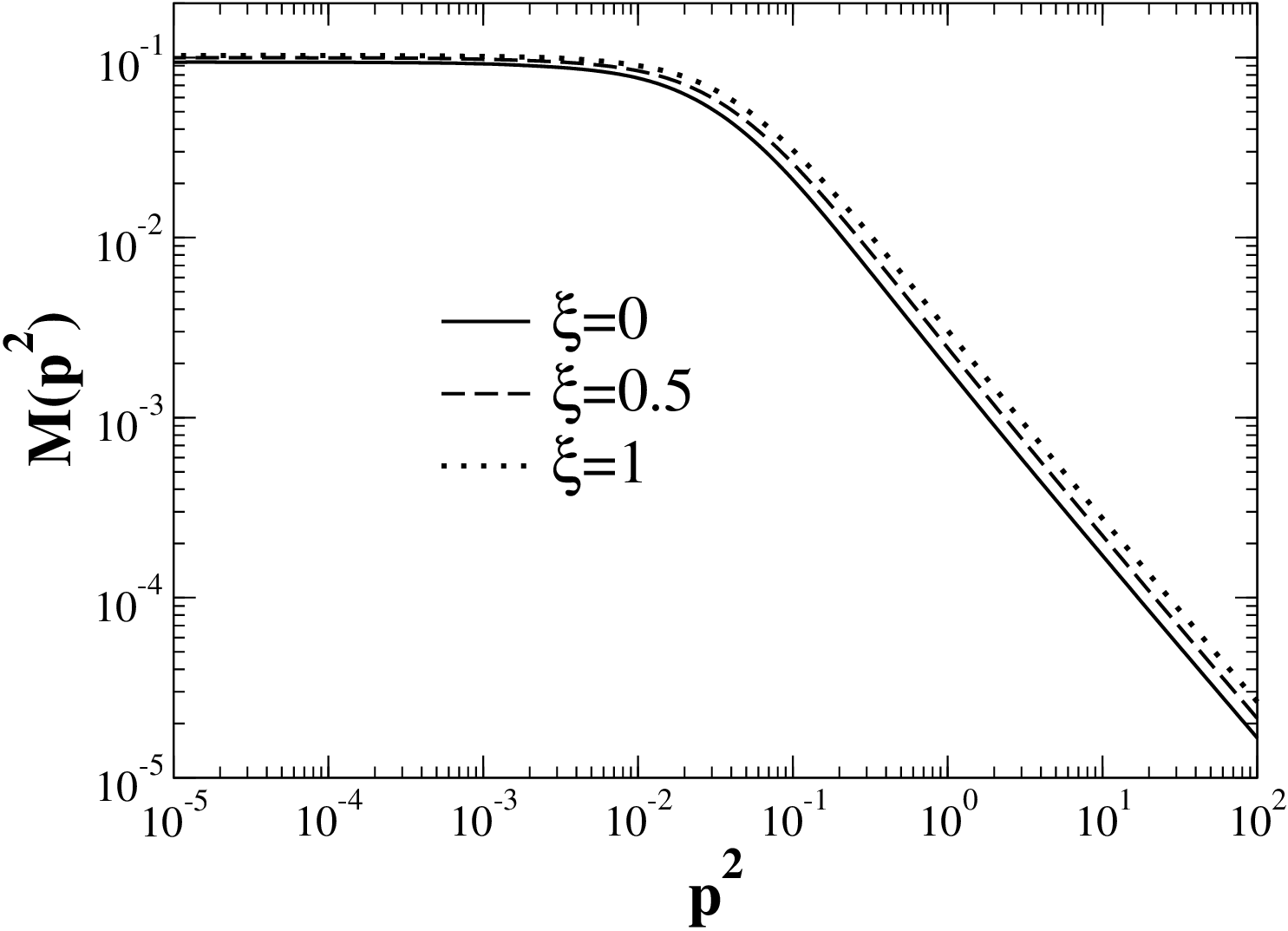,width=7.5cm,height=6cm}
}\parbox{7.5cm}{\epsfig{file=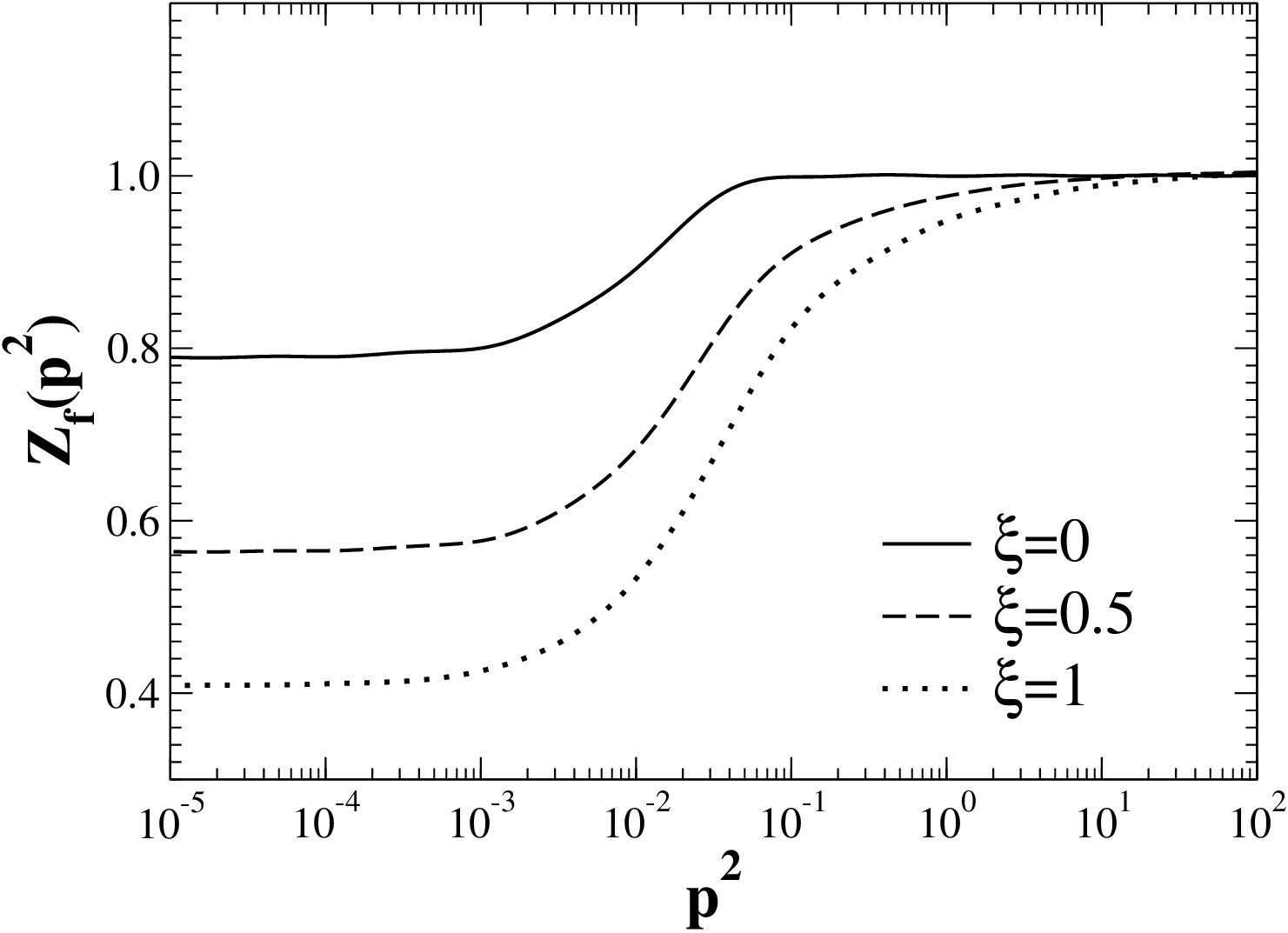,width=7.5cm,height=6cm}}
\caption{\label{fig1} 
Here we display the fermion mass function, $M(p^2)$, and  the wave function
renormalisation, $Z_f(p^2) = 1/A(p^2)$ in quenched approximation for 
three different values of the
gauge parameter $\xi$.  The scale is set by choosing $e^2=1$.}
\vspace{0.5cm}
\centerline{
\epsfig{file=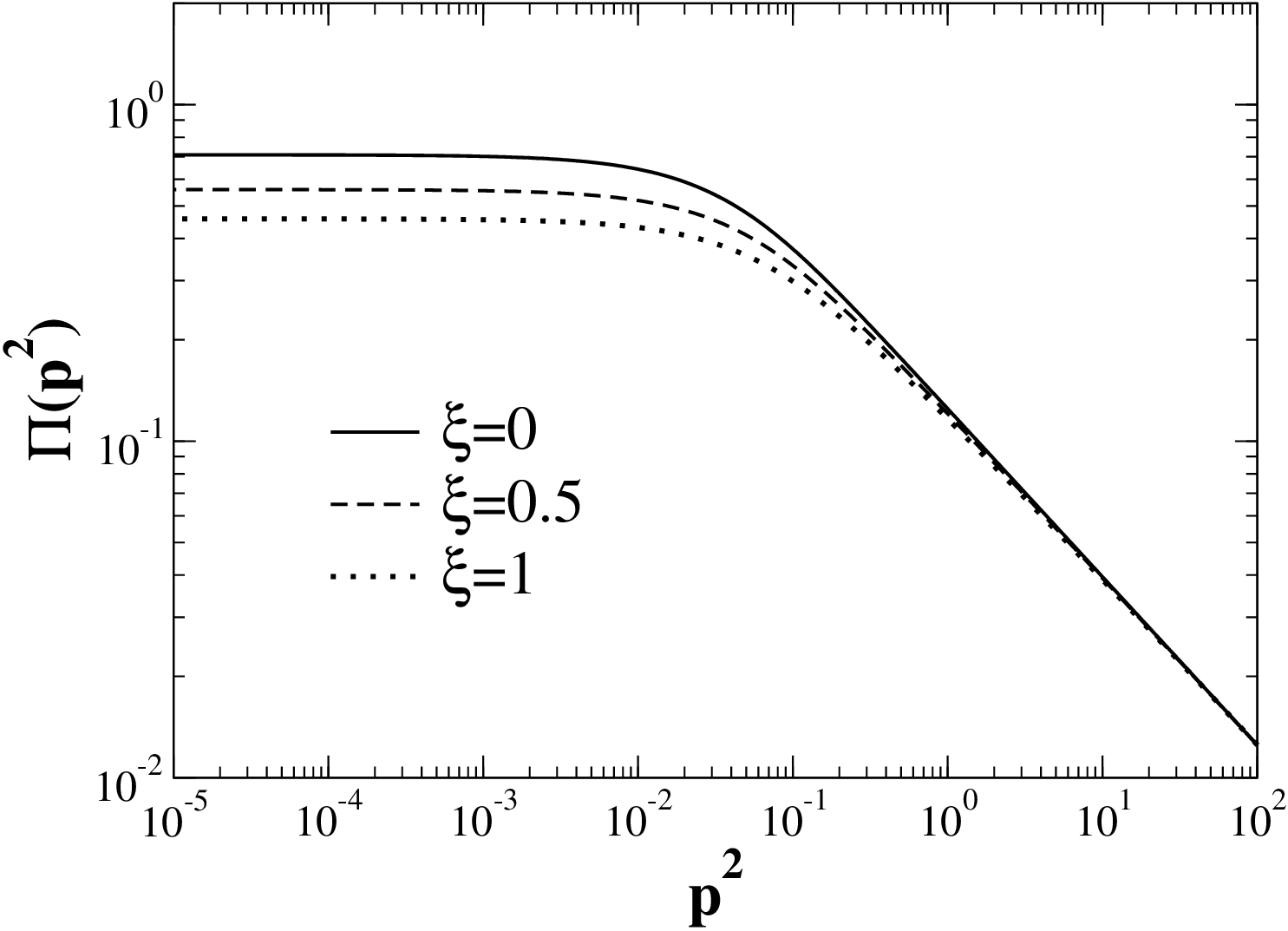,width=7.5cm,height=6cm}
}
\caption{\label{fig1a} 
Here we display the photon polarisation
$\Pi(p^2)$ as calculated from the photon DSE
using $N_f=1$ and the quenched fermion propagator functions 
without back-coupling
for three different values of the
gauge parameter $\xi$.  The scale is set by choosing $e^2=1$.}
\end{figure}

The corresponding numerical solutions are displayed in
Figs.~\ref{fig1} and \ref{fig1a}.  For the fermion propagator, no
qualitative difference between Landau gauge and gauges with
non-vanishing gauge parameter is found.  Notice that $0 < Z_f(p^2) \le
1$ on the entire momentum range, as one would expect from quenched
perturbative theory, eq.(\ref{UV_A}).  For the photon polarisation we
find sizable violations of gauge invariance employing the BC-vertex in
accordance with Ref.~\cite{Burden:1991uh}.  Adding the CP term also in
the photon DSE does not help in this respect and furthermore
introduces spurious divergences.

\end{appendix}

\pagebreak


\end{document}